\documentclass[12pt]{article}
\usepackage{amsmath,amssymb,mathtools,nccmath}
\usepackage{bm,bbm}
\usepackage{graphicx,psfrag,epsf,esvect}
\usepackage{enumerate}
\usepackage{natbib}
\usepackage{url} 
\usepackage[utf8]{inputenc}
\usepackage[ruled]{algorithm2e}
\usepackage[font=footnotesize]{caption}
\usepackage{subcaption}
\usepackage[toc,page,titletoc]{appendix}
\usepackage{xcolor}
\usepackage{float}
\usepackage{setspace}
\usepackage{multirow}

\makeatletter
\g@addto@macro{\normalsize}{%
   \setlength{\abovedisplayskip}{5pt}
   \setlength{\abovedisplayshortskip}{5pt}
   \setlength{\belowdisplayskip}{5pt}
   \setlength{\belowdisplayshortskip}{5pt}}
\makeatother

\usepackage{authblk}

\title{Stochastic Variational Methods in Generalized Hidden Semi-Markov Models to Characterize Functionality in Random Heteropolymers} 
\author[1]{Yun Zhou}
\author[1]{Boying Gong}
\author[2]{Tao Jiang}
\author[3,4,5]{Ting Xu}
\author[6,7]{Haiyan Huang\thanks{Corresponding author. Email: hyh0110@berkeley.edu}}
\affil[1]{Division of Biostatistics, University of California Berkeley, Berkeley, CA, USA.}
\affil[2]{Department of Chemistry, The MOE Key Laboratory of Spectrochemical Analysis and Instrumentation, Xiamen University, Xiamen, China.}
\affil[3]{Department of Materials Science and Engineering, University of California Berkeley, Berkeley, CA, USA. }
\affil[4]{Department of Chemistry, University of California Berkeley, Berkeley, CA, USA.}
\affil[5]{Materials Science Division, Lawrence Berkeley National Laboratory, Berkeley, CA, USA.}
\affil[6]{Department of Statistics, University of California Berkeley, Berkeley, CA, USA.}
\affil[7]{Center for Computational Biology, University of California Berkeley, Berkeley, CA, USA.}

\date{}

\bibpunct[, ]{(}{)}{;}{a}{,}{,}

\begin{document}

\def\spacingset#1{\renewcommand{\baselinestretch}%
{#1}\small\normalsize} \spacingset{1}
%

 \maketitle

\bigskip
\begin{abstract}

Recent years have seen substantial advances in the development of biofunctional materials using synthetic polymers.
The growing problem of elusive sequence-functionality relations for most biomaterials has driven researchers to seek more effective tools and analysis methods.
In this study, statistical models are used to study sequence features of the recently reported random heteropolymers (RHP), which transport protons across lipid bilayers selectively and rapidly like natural proton channels. We utilized the probabilistic graphical model framework and developed a generalized hidden semi-Markov model (GHSMM-RHP)
to extract the function-determining sequence features, including the transmembrane segments within a chain and the sequence heterogeneity among different chains.
We developed stochastic variational methods
for efficient inference on parameter estimation and predictions,
and empirically studied their computational performance 
from a comparative perspective on Bayesian (i.e., stochastic variational Bayes) versus frequentist (i.e., stochastic variational expectation-maximization) frameworks that have been studied separately before.
The real data results agree well with the laboratory experiments, 
and suggest GHSMM-RHP's potential in predicting
protein-like behavior at the polymer-chain level.

\end{abstract}

\vfill

\newpage
\spacingset{1} 

\section{Introduction}
\label{sec:intro}

Protein-mimicry with synthetic polymers has been actively pursued for decades to meet material demands within the environmental, energy and life sciences. Recently, \cite{panganiban2018random} and \cite{jiang2020single} reported a promising protein-mimic polymer system on the basis of random heterogeneous polymers (RHP) comprising monomers with distinct chemical properties. Through wet lab experiments, they found long-term screening for optimizing the monomer choices and monomer ratios can lead to some RHPs with protein-like functions. 
The RHP systems provide an attractive platform for developing polymer-based biofunctional materials.

In \cite{jiang2020single},
several datasets of four-monomer based RHPs were synthesized, which behave like natural protein-based proton channels. The four methacrylate-based monomers were selected by chemical diversity, i.e., methyl methacrylate (MMA), 2-ethylhexyl methacrylate (EHMA), oligo (ethylene glycol) methacrylate (OEGMA), and 3-sulfopropyl methacrylate potassium salt (SPMA). MMA and EHMA are the hydrophobic monomers, while OEGMA and SPMA are the hydrophilic monomers.
RAFT polymerization was used to control the ratio of four monomers in an RHP dataset, in order to tailor the overall hydrophobicity.
Different RHP datasets with varying monomer ratios were generated, and some of them were found capable of being inserted into lipid bilayers and promoting transmembrane proton transport.
However, there is a difference from natural proteins that make up the channels. A functional natural protein has certain specific amino acid sequence(s). But the functional performance of each RHP dataset is determined as a whole, and chains in an RHP dataset, which are synthesized under a fixed monomer ratio, are largely different from each other.
The sequence heterogeneity in RHP brings challenges to directly linking individual monomer sequence to the overall performance of a polymer dataset.
Study of the sequence heterogeneity under various monomer composition ratios would shed light on the understanding of RHPs' functional adaptability, enabling their uniform behavior in various environments. More importantly, it offers promising potential for further RHP design in a predictable manner.

As RHPs' local hydrophobicity levels relate to short-range interactions, 
each RHP chain can be studied as a concatenation of segments of various cumulative hydrophobicity.
It is believed that varied segment lengths and ending positions substantially affect RHPs' chemical properties.
Therefore, segmentation can reveal valuable latent structure information on RHP heterogeneity and functionalities when individual chain experimentation is unavailable.
The differences in cumulative hydrophobicity among segments introduce the segment heterogeneity.
To analyze such heterogeneity,
the segments are assigned to three states: the hydrophilic state, $S1$, for the segments rich in hydrophilic monomer OEGMA and the negatively charged SPMA; the hydrophobic state, $S3$, for those comprising mostly hydrophobic monomers MMA and EHMA; the intermediate state, $S2$, for ones that cannot be simply assigned as hydrophobic or hydrophilic (amphiphilic segments). The $S3$ segments have the ability to span across the bilayer region and provide the proton transport pathway. The $S1$ segments prefer the water environment. The $S2$ segments are believed to have short residence time in both bilayer and water regions.

\begin{figure}
\centering
     \begin{subfigure}[b]{0.35\textwidth}
         \centering
         \includegraphics[width=\textwidth]{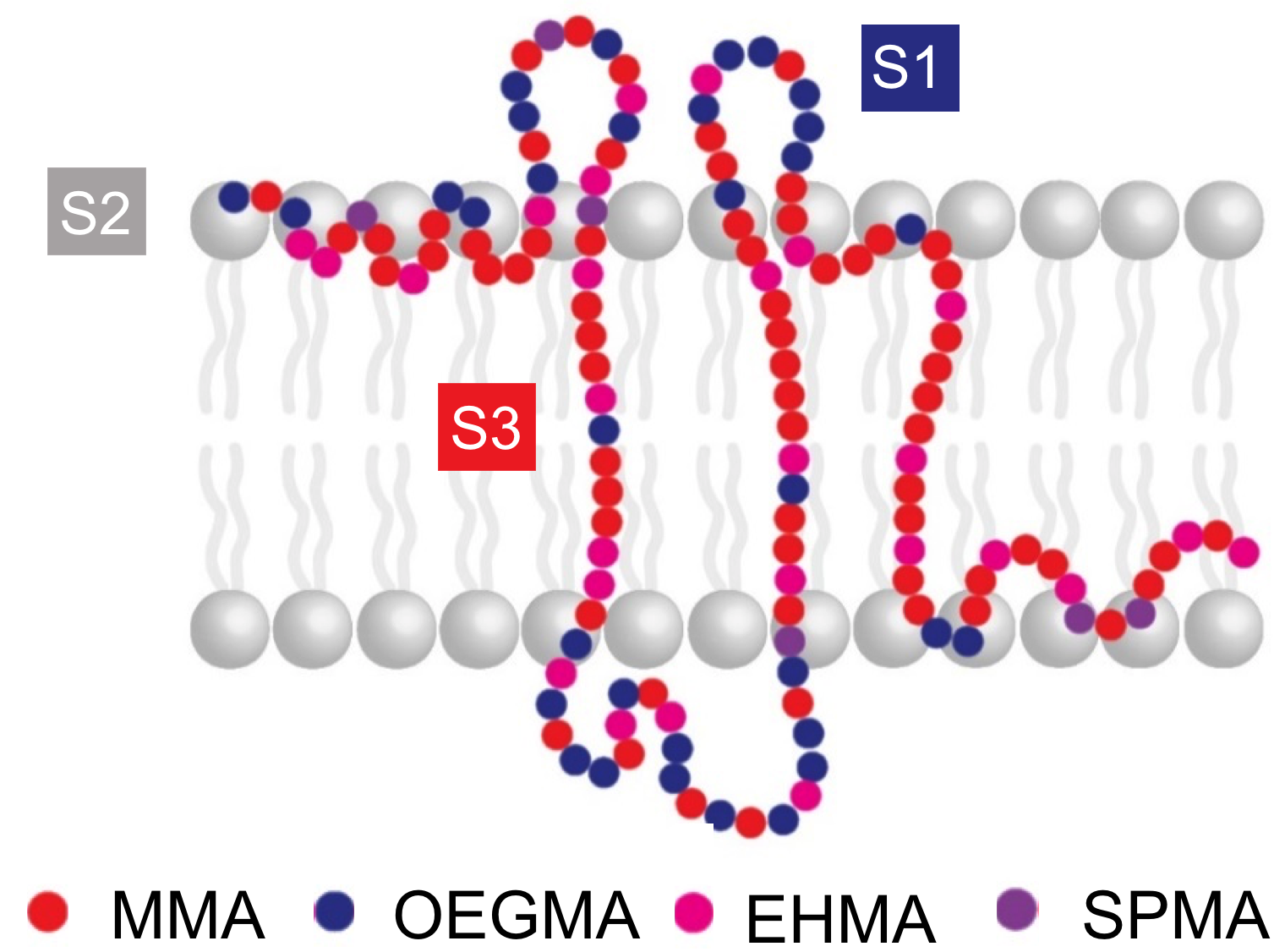}
         \caption{\textbf{Two dimensional.}}
     \end{subfigure}
     \begin{subfigure}[b]{0.35\textwidth}
         \centering
         \includegraphics[width=\textwidth]{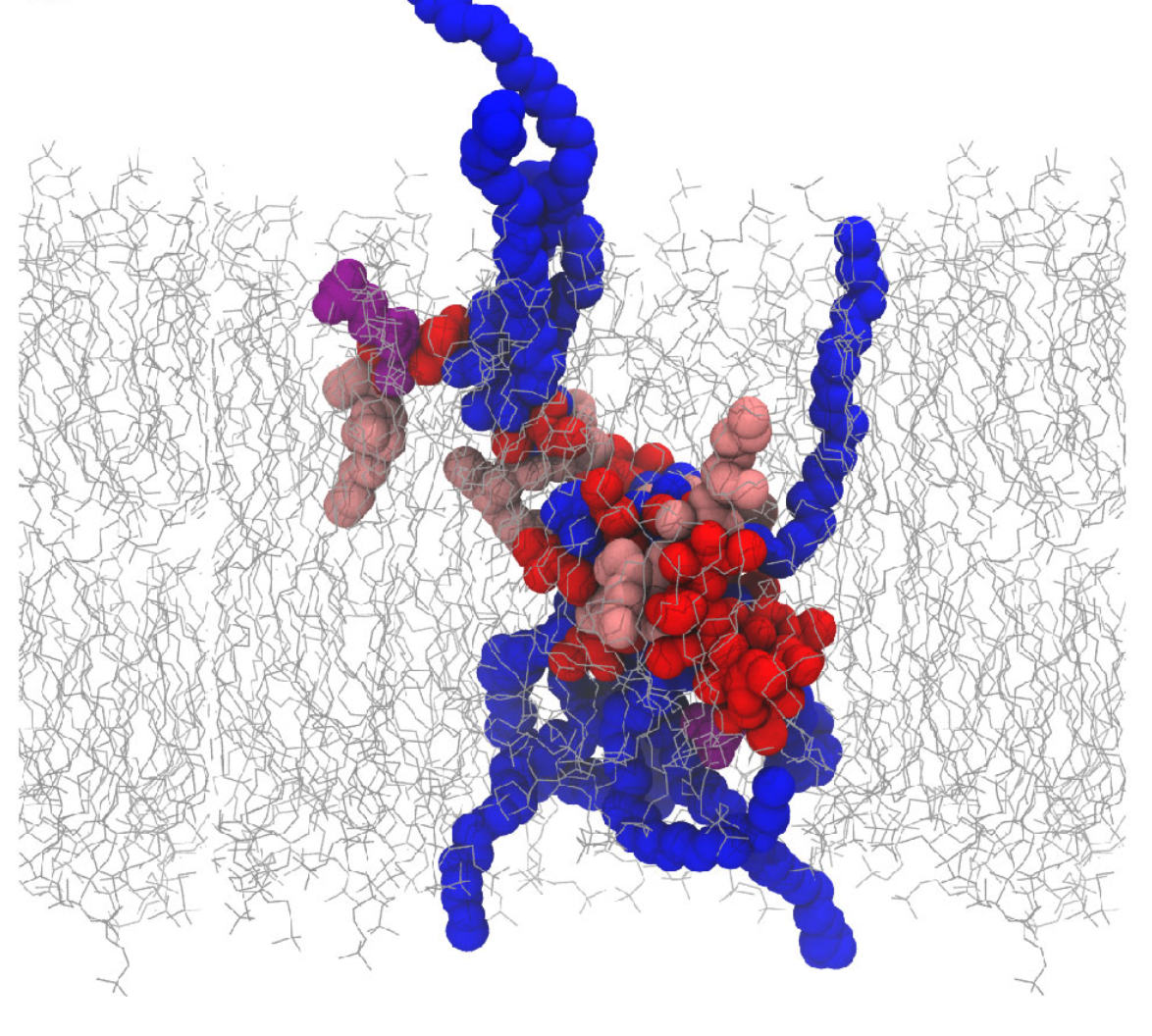}
         \caption{\textbf{Three dimensional.}}
     \end{subfigure}
\caption{\textbf{Schematic illustration of RHP heterogeneity}. (a): 2D illustration of one RHP chain in a lipid bilayer. (b): Spatial 3D illustration of the RHP in the lipid bilayer. MMA, OEGMA, EHMA, and SPMA are shown in colors of red, blue, pink, and purple, respectively. Lipid is shown in grey color.
The figures are adapted from \cite{jiang2020single}.
Note that the 2D and 3D structures are shown for schematic illustration only. In practice we only have 1D information, i.e., the linear monomer sequence in each RHP dataset.}
\label{fig:BG}
\end{figure}

Figure \ref{fig:BG} displays the schematic illustration of RHP chains in a lipid bilayer. Note however, information on segmentation and spatial structure is not available in practice, and we only have access to the linear sequence of four monomers in each RHP dataset.

Determining how RHPs' performance relates to segment heterogeneity and sequence heterogeneity, rather than to a specific realization of polymer sequence, provides a promising strategy for developing biofunctional materials.
The two types of heterogeneity correspond to two key factors in RHP design: (1) the segmentation of local hydrophobicity that regulates short-range interactions, and (2) the choice of overall monomer composition ratio that modulates the solubility.
So far, domain knowledge is extremely limited in the field. There is also a lack of statistical study that could bridge the experimental restrictions and RHPs' predictable performance.
To exploit the statistical design of RHP and to aid its synthesis, we utilized the probabilistic graphical model framework and developed a generalized hidden semi-Markov model (GHSMM-RHP).
Based on this model, we predicted the segments of varying hydrophobicity within RHP chains to address the first key factor above. To address the second factor, we statistically quantified the heterogeneity of chain distributions among different RHP datasets.
GHSMM-RHP helps to extract the hidden phenomena that are otherwise hard to quantify, and utilizes efficient approximate inference in place of expensive exact computation.
That is, we aim to address a joint estimation-prediction problem in RHP systems under the GHSMM-RHP model.

Specifically,
GHSMM-RHP adopts the residential time design, which has fewer parameters and lower time complexity than generic hidden semi-Markov model \citep{yu2010hidden}.
GHSMM-RHP incorporates additional layer of nodes to allow handling segment constraints such as the monomer positions, as outlined in Section \ref{sec:Model}. 
Based on the model, we developed stochastic variational methods in closed-form under both frequentist and Bayesian settings.
To keep the monomer dependencies within an RHP chain, 
we derived message-passing formulas on the graphical model of conjugate exponential families \citep{maathuis2018handbook}. 
GHSMM-RHP differs from existing efforts in its novel design that incorporates specific constraints for large RHP systems,
and provides a scalable graphical model with efficient algorithms that meets the inference demand in RHP applications.

In addition, we studied the GHSMM-RHP model through a simulation analysis, 
and compared estimates from the frequentist and Bayesian stochastic variational methods. 
In particular,
we implemented stochastic variational expectation-maximization (SVEM) and stochastic variational Bayesian (SVB), and empirically addressed their estimation/prediction accuracy and computational (in)stability against hyper-parameters.
We found that both methods achieved low statistical error, while SVEM exhibited greater stability against random initializations.
Our work makes the first attempt to 
bring the two stochastic variational methods, which were studied separately before, to the same nonconvex regime. It also contributes to the current literature on the computational error of stochastic variational methods.

We further evaluated the GHSMM-RHP model using real datasets.
\cite{jiang2020single} synthesized seven RHP datasets with varied compositional ratios and tested their performance. 
Among them, RHP with a compositional ratio of 50(MMA) : 20(EHMA) : 25(OEGMA) : 5(SPMA) was the most functionally active, which we refer to as RHP50/20.
We used RHP50/20 as a reference and compared other RHPs to RHP50/20 based on the GHSMM-RHP model.
The RHP datasets identified by our model as having high similarity to RHP50/20 all produced high proton transport efficiency in the lab experiment, which
suggests that the GHSMM-RHP model has the potential to aid in predictable RHP performance.
By bridging the gap between costly laboratory assessments and less expensive computational models, we hope to provide guidance for designing improved RHP syntheses.

The rest of the paper is organized as follows. 
In Section \ref{sec:relevant} we review the background of methods related to the GHSMM-RHP model.
In Section \ref{sec:Model} we describe the model that incorporates the constraints of RHP systems. In Section \ref{sec:Method} different stochastic variational methods are derived. We implemented the algorithms and compared the results in Section \ref{sec:Result}. Section \ref{sec:Conc} concludes the work and discusses avenues for further research.

\section{Method Background Review}
\label{sec:relevant}

\subsection{Relevant Work on the GHSMM-RHP Model}

The GHSMM-RHP model is inspired by previous attempts by \cite{jiang2020single} to use a Hidden Markov model (HMM) to analyze certain restricted RHP datasets.
HMMs are widely used to describe a sequence of observable events which are generated by a sequence of internal hidden states. It has a long successful history in biological sequence analysis, such as gene prediction \citep{munch2006automatic} and modeling DNA sequencing errors \citep{lottaz2003modeling}.

We noticed two drawbacks of using HMM in RHP applications.
First, the hidden state length distribution in HMM implicitly follows a geometric distribution. 
However, the segment length of RHP chains has a clear lower and upper cutoff, which needs to be configurable in the model to test key polymerization assumptions \citep{jiang2020single}.
As suggested by \cite{rabiner1989tutorial}, more flexible state durations need to be incorporated to improve parameter estimation and segment prediction.
To overcome this issue we adopted a 
semi-Markov modeling of the underlying hidden states \citep{yu2010hidden}. Such a duration model has a more scalable graphical architecture to account for segmental heterogeneity. 
The other downside involves the time complexity for inference algorithms. The widely used Baum-Welch algorithm \citep{baum1970maximization} takes $TNKS^2$ time under a classic HMM,
and $TNKS(S+D)$ under residential time model (a semi-Markov model described in \cite{yu2003efficient}),
where T is the number of iterations, N is the total number of sequences, K is the number of monomers in one RHP chain, S is the number of hidden states,
and D is the span of the hidden state length distribution.
The algorithms become time-consuming when training large libraries of RHP datasets. 
This computational issue can be
alleviated by implementing stochastic variational methods, where stochastic optimizations are enabled through processing only a small portion of the entire dataset in each iteration \citep{robbins1951stochastic, bottou2010large}.

The basic idea of variational methods used in GHSMM-RHP is to turn statistical estimation into a simpler lower bound optimization problem with tractable approximation densities. 
It defines a family of distributions $\mathcal{Q}$ over the latent variables (i.e., the hidden state sequences underlying the RHP chain), each considered as a candidate approximation to the true posterior or conditional probabilities. 
$\mathcal{Q}$ should be large enough to reduce the approximation error, but also simple enough to allow for efficient optimization algorithms.
Among the related work,
\cite{ghahramani2001propagation} and \cite{winn2005variational} connected the junction tree algorithms on graphical models with the variational Bayesian posterior distributions.
\cite{mackay1997ensemble} and \cite{beal2003variational} were among the first efforts to implement variational Bayesian on classic HMM models. 
\cite{hoffman2013stochastic} introduced stochastic optimization into variational Bayesian, 
and extensions to other Bayesian time series models were reported in \cite{johnson2014stochastic}. 
\cite{foti2014stochastic} used stochastic variational Bayesian for HMM but in a different setting: it involved a single long sequence broken into mini-batches, rather than multiple relatively short sequences.

\subsection{Relevant Work on Stochastic Variational Methods}

Variational methods, such as variational expectation-maximization (VEM) and variational Bayes (VB, sometimes termed variational inference), provide alternative solutions to maximum likelihood estimation and Bayesian posterior approximation respectively \citep{wainwright2008graphical, blei2017variational}. 
Being faster than convex relaxations and Markov chain Monte Carlo as well as adaptable to different application models, variational methods have prospered in various large scale approximate inference problems such as computational biology \citep{carbonetto2012scalable}, natural language processing \citep{yogatama2014dynamic}, and astronomy \citep{regier2015celeste}. 
The objective function by classic proposals \citep{neal1998view, jordan1999introduction} is:
\begin{align}
\label{variational}
 \ln p_{\theta}(y) 
 = \ln \int p_{\theta}(y, z) dz  
 \geq  \ln p_{\theta}(y)  -  \text{KL}(q(z) || p_{\theta}(z | y) ). 
\end{align}
Here, the likelihood function $ p_{\theta}(y)$ can be replaced by marginal $p(y)$ in the Bayesian setting, where $z$ are latent variables that include parameters $\theta$. 
One can choose $q(z)$ over some proper variational distribution family $\mathcal{Q}$.
Based on Equation \ref{variational}, some variants include
those that are scaled up
by stochastic approximations \citep{cappe2009line, hoffman2013stochastic},
and those that are distinct in variational objective formulations (e.g., $\alpha$ -VB \citep{yang2020alpha}) 
or distribution families $\mathcal{Q}$ (e.g., Auto-encoding VB \citep{kingma2013auto}, and Boosting VB \citep{guo2016boosting}).

Most theoretical justifications for variational methods, which are still in active development, focus on the statistical properties of the global optimizer (either point estimates or posterior approximations). 
For example, 
the consistency and asymptotic normality in
stochastic block models \citep{bickel2013asymptotic},
mixture models \citep{westling2019beyond},
or more general parametric models with mean-field assumptions \citep{wang2019frequentist};
the non-asymptotic convergence rate for
latent Gaussian models \citep{sheth2017excess},
approximating tempered posterior \citep{alquier2020concentration},
point estimates using mean-field assumptions \citep{pati2018statistical},
or high-dimensional and nonparametric inference \citep{zhang2020convergence}.

Although the global optimum for variational methods can be achieved via some suitable settings \citep{awasthi2015some},
the computational error involving the convergence behavior of iterative updates has been less studied. 
In fact, solutions to variational objective functions often suffer from multiple local optimums, making optimizations sensitive to initializations and other hyper-parameters.
For example, unfavorable scenarios were constructed in the stochastic block model with futile random initializations \citep{mukherjee2018mean}, and in the topic model with uninformative objective functions \citep{ghorbani2019instability}. 
They point out cases that fail in practice due to either optimization algorithms or problem formulations.

A few studies have investigated scenarios
in which the global optimum can not be attained feasibly. For example, \cite{zhang2020theoretical} analyzed the computational rate of convergence for mean-field variational Bayes in stochastic block models, and \cite{balakrishnan2017statistical} studied EM-type of iterates in broader frequentist settings. 
However, those results are either model-specific or conditional on proper hyper-parameters, and the additional tuning effort with pilot algorithms may not be warranted.
Improvement of the computational error in variational methods with general initializations and model choice remains an open problem. Moreover, when applying variational methods in a broad range of model types, poor approximations can originate from the instability of stochastic optimization algorithms with many hyper-parameters, even if the variational family $\mathcal{Q}$ contains the true posterior \citep{dhaka2020robust}.

Results on the accuracy and stability of stochastic variational methods remain inconclusive for a wide variety of models. Misperception of its properties will hinder reproducible scientific research and reliable decision making.
In this paper, we make an attempt to 
empirically study the computational performance of stochastic variational methods under the GHSMM-RHP model, 
as well as to compare the performance of stochastic variation methods under Bayesian (i.e., SVB) and frequentist (i.e., SVEM) frameworks that were previously studied separately. 
The observations and conclusions developed in our study from a comparative perspective on SVB and SVEM
not only provide insight into the choice of these two methods,
but also contribute to the existing literature on the computational error of stochastic variational methods.

\section{Generalized Hidden Semi-Markov Model for RHP}
\label{sec:Model}

As described previously, each RHP dataset contains thousands of sequences randomly generated under a fixed monomer ratio.
Note that we only have access to the one-dimensional primary structure, that is, the linear sequence of four monomers.
We start by discussing a few sequence constraints, which are the same as those stated in \cite{jiang2020single}.
These constraints describe the underlying biology of functional RHPs: (1) We define $S3$ segments to be composed of hydrophobic monomers (i.e., MMA, EHMA) and no more than one OEGMA. If an OEGMA appears in an S3 segment, it must be at least two monomers away from both ends of the $S3$ segment. (2) $S2$ and $S3$ are separated by hydrophilic segments $S1$. (3) The negatively charged hydrophilic monomer SPMA appears only in $S1$.

The hydrophobic monomers in $S3$ segments enable the insertion of the chain into lipid bilayers, and the appearance of a single OEGMA doesn't compromise the overall activity if it is in the middle of an $S3$ segment.
The length of the $S3$ segments depends on various factors such as chain conformation and membrane flexibility, but it should be sufficiently long to span the core of the lipid bilayer.
The $S1$ segments have a high preference for water (high polarity of the monomer), and appear in between $S2$ and $S3$.

We adapted the right-censored hidden semi-Markov model in \citep{yu2010hidden} to model the RHP chain while incorporating the above constraints. 
Here, the hidden state duration is a random variable following a categorical distribution over a predefined range.
The unobserved states correspond to three types of segments: hydrophilic state $S1$, intermediate state $S2$, and hydrophobic state $S3$. We denote a random RHP chain as $(y_1, ..., y_K)$. At each position $k \in 1, ..., K$, $y_k$ takes values from the four monomers, i.e., MMA, OEGMA, EHMA, SPMA.
$(z_k, \tau_k, \iota_k)$ are the latent variables. $z_k$ is the hidden segment state, $\tau_k$ is the remaining duration of the current hidden state, and $\iota_k$ is the location of an emitted OEGMA. 
Their distributions are described below.

$\bullet \hspace{1mm} z_1$:
The model starts with the initial state variable $z_1$, taking value in the state space $\mathcal{S}_z = \{ S1, S2, S3\}$. 
$p(z_1 | \pi)$ follows a categorical distribution with parameters $\pi  = (\pi_{S1}, \pi_{S2}, \pi_{S3})$:
\begin{equation}
\begin{aligned}
\label{eq:initial}
\ln p(z_1 | \pi) = 
\sum_{i \in \mathcal{S}_z} \ln(\pi_i) \mathbbm{1} \{z_1 = i\}. 
\end{aligned}
\end{equation}

$\bullet \hspace{1mm} (\tau_1, \iota_1)$:
Latent variables $(\tau_1, \iota_1)$ depend on $z_1$. 
(1) If $z_1 = S3$, then the duration $\tau_1$ takes values between $\underline{D}$ and $\overline{D}$, the predefined shortest and longest segment length. 
The OEGMA location $\iota_1$ takes values between 3 and $\tau_1 - 2$, meaning the OEGMA shall be at least two monomers away from both ends of the $S3$ segment. $\iota_1$ may also be 0, meaning there is no OEGMA in the current $S3$ segment. 
This support is denoted as $\mathcal{S}^{S3}_{\tau,\iota} = \{ (d,l) | (d,l) \in \{\underline{D},...,\overline{D}\} \times \{0, 3,...,\overline{D}-2\}, l \leq d-2\}$.
The conditional probability $p(\tau_1, \iota_1 | z_1, \rho^{S3})$ follows a categorical distribution with parameters $\rho^{S3} = \{  \rho^{S3}_{d,l} | (d,l) \in \mathcal{S}^{S3}_{\tau,\iota}   \}$. 
(2) If $z_1 = S1$ or $z_1 = S2$, then the duration $\tau_1$ takes value between $1$ and $\overline{D}$. 
There is no constraint on OEGMA location for $S1$ and $S2$ segments, so $\iota_1$ can take any value without affecting the likelihood function. We set it to 0 for convenience.
This support is denoted as $\mathcal{S}^{S1}_{\tau,\iota} = \mathcal{S}^{S2}_{\tau,\iota}  = \{ (d,l) | (d,l) \in \{1,...,\overline{D}\} \times \{0\} \}$.
The conditional probability $p(\tau_1, \iota_1 | z_1, \rho^{S1})$ follows a categorical distribution with parameters $\rho^{S1} = \{  \rho^{S1}_{d,l} | (d,l) \in \mathcal{S}^{S1}_{\tau,\iota}   \}$. $p(\tau_1, \iota_1 | z_1, \rho^{S2})$ and $\rho^{S2}$ is defined similarly.

In summary, the conditional probability $p(\tau_1, \iota_1 | z_1, \rho)$ follows a categorical distribution with parameters $\rho = (\rho^{S1}, \rho^{S2}, \rho^{S3})$. 
Three sub-vectors $\rho^{S1}, \rho^{S2}, \rho^{S3}$ correspond to three supports $\mathcal{S}^{S1}_{\tau,\iota}, \mathcal{S}^{S2}_{\tau,\iota}, \mathcal{S}^{S3}_{\tau,\iota}$ respectively. This gives:
\begin{equation}
\begin{aligned}
\label{eq:duration}
\ln p(\tau_1, \iota_1 |z_1, \rho) =  
\sum_{i \in \mathcal{S}_z} \sum_{(d,l) \in \mathcal{S}^i_{\tau, \iota}}\ln(\rho^{i}_{d,l}) \mathbbm{1} \{(z_1, \tau_1, \iota_1) = (i,d,l)\}.
\end{aligned}
\end{equation}

\begin{figure}
\begin{center}
\includegraphics[width=2.5in]{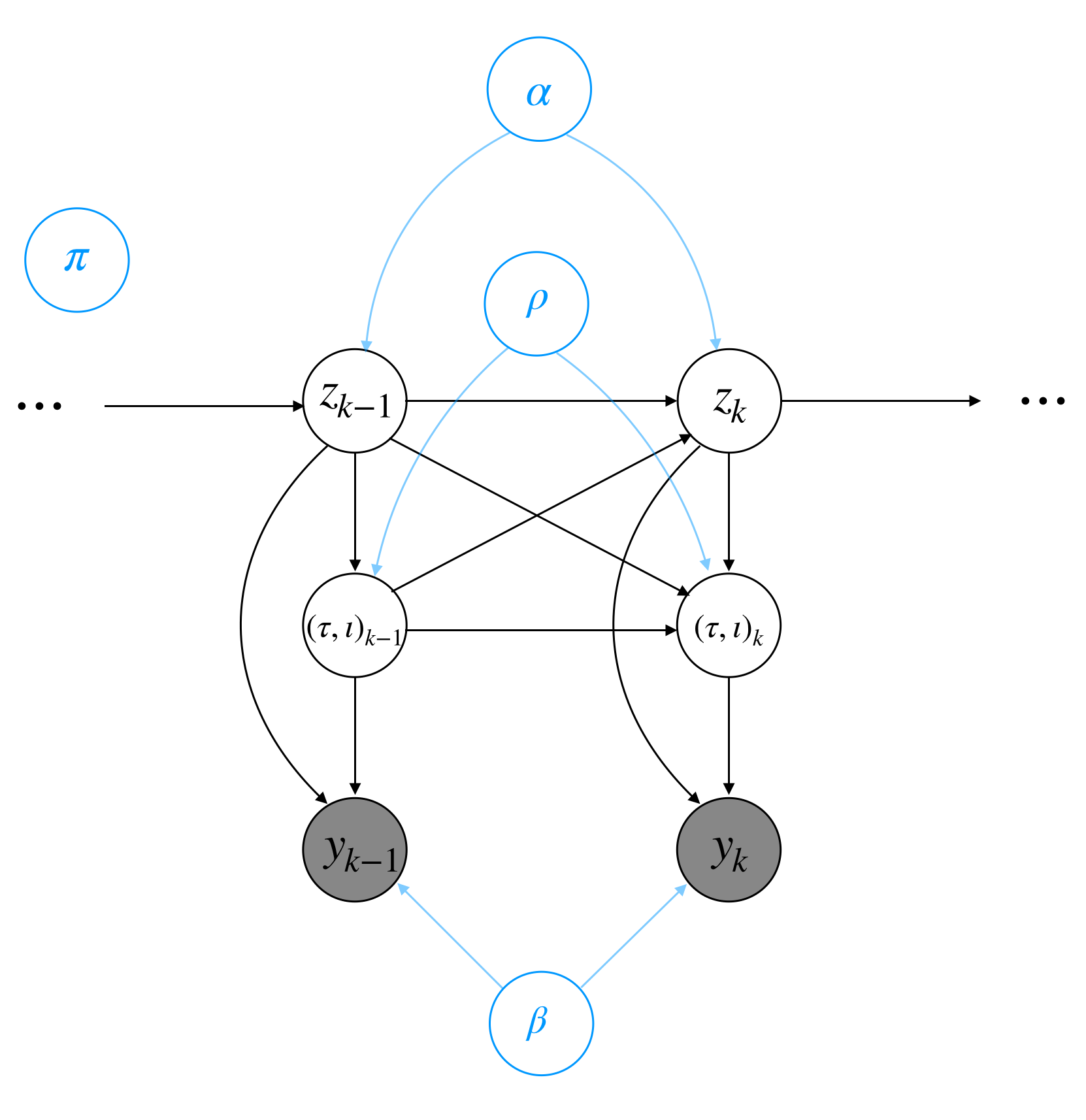}
\end{center}
\caption{Illustration of the graphical model in GHSMM-RHP.}
\label{fig:HSMM_PGM}
\end{figure}

$\bullet \hspace{1mm} (z_k, \tau_k, \iota_k)$:
For positions $k = 2, ..., K$, latent variables $(z_k, \tau_k, \iota_k)$ evolve as a time-homogeneous Markov chain, taking values in the same joint space as the one described for $(z_1, \tau_1, \iota_1)$.
The graphical model diagram is illustrated in Figure \ref{fig:HSMM_PGM}.
The transition distribution with parameters $\alpha, \rho$ is defined as $\zeta_{ (z_{k-1}, \tau_{k-1}, \iota_{k-1}) \to (z_{k}, \tau_{k}, \iota_{k}) }$.
When $(z_{k}, \tau_{k}, \iota_{k}) = (j,d',l')$ and $ (z_{k-1}, \tau_{k-1}, \iota_{k-1}) = (i, d, l)$,
the transition kernel $\zeta$ is:
\begin{equation*}
\begin{aligned}
  \zeta_{(i,d,l) \to (j,d',l')} =
  \begin{cases}
    1, & \text{if $d>1, $}  \text{$(j,d',l') = (i,d-1,l)$}.  \\
    \alpha_{i,j}\rho^{j}_{d', l'}, & \text{if $d=1, j \neq i $}.\\
    0, & \text{otherwise}.
  \end{cases}
\end{aligned}
\end{equation*}

As the position $k$ proceeds we note the following. 
(1) If $d > 1$, 
there is no transition on hidden segment state, i.e., $z_k = j = i$.
The remaining duration $\tau_{k-1} = d$ of hidden segment state $z_{k-1} = i$ will deterministically decrease by 1 until it reaches 1, i.e., $\tau_k = d - 1$, $\tau_{k+1} = d - 2$, et cetera.
(2) If $ d = 1$,
the RHP will transit to a different hidden segment state $z_k = j \in \mathcal{S}_z  \setminus \{ i \} $ with probability $\alpha_{i,j}$, followed by a new pair $(\tau_k, \iota_k) = (d', l')$ generated according to the categorical probability parameter $\rho^{j}$.
We disallow transitions between S2 and S3 due to biological constraints, i.e., $\alpha_{S2,S3} = \alpha_{S3,S2} = 0$. Thus, we have:
\begin{equation}
\begin{aligned}
\label{eq:transition}
&\ln 
[p( z_k, \tau_k, \iota_k |  z_{k-1}, \tau_{k-1}, \iota_{k-1},   \alpha, \rho) 
\\
= & \sum_{i, j\in \mathcal{S}_z} \sum_{(d,l) \in \mathcal{S}^i_{\tau, \iota}} \sum_{(d',l') \in \mathcal{S}^j_{\tau, \iota}} \Big[ \ln(\zeta_{(i,d,l) \to (j,d',l')}) \cdot 
\mathbbm{1} \{ (z_{k-1}, \tau_{k-1}, \iota_{k-1}) = (i, d, l), (z_k, \tau_k, \iota_k) = (j, d', l')\} \Big]. 
\end{aligned}
\end{equation}

$\bullet \hspace{1mm} y_k$:
The observed random variable 
$y_k$ is conditionally independent of all the other variables given $(z_k, \tau_k, \iota_k)$. $y_k$ takes values in the space of four monomers \linebreak $\mathcal{S}_y = \{MMA, OEGMA, EHMA, SPMA\}$.
The emission distribution $p(y_k | z_k, \tau_k, \iota_k, \beta)$ follows a categorical distribution with parameters 
$\beta = (\beta_{S1}, \beta_{S2}, \beta_{S3}, \beta_{S3'})$.
The four sub-vectors of $\beta$ are used conditional on different values of $(z_k, \tau_k, \iota_k)$,
as shown in Table \ref{tab:emis_prob_para}.

\begin{table}
\caption{Parameters for emission distribution.}
\label{tab:emis_prob_para}
\begin{center}
\begin{tabular}{rrr}
Emission distribution  & Support of $y_k$ \\\hline
$p(y_k = i | z_k = S1, \beta) = \beta_{S1,i} $ &   $i \in \mathcal{S}_y$  \\
$p(y_k = i | z_k = S2, \beta) = \beta_{S2,i} $  &  $i \in \mathcal{S}_y  \setminus \{SPMA\}$  \\
$p(y_k = i | z_k = S3, \tau_k \neq \iota_k  \text{ or }  z_k = S3, \iota_k = 0, \beta) = \beta_{S3,i} $  & $i \in \mathcal{S}_y  \setminus \{OEGMA, SPMA\}$  \\
$p(y_k = i | z_k = S3, \tau_k=\iota_k, \beta) = \beta_{S3',i}  = 1 $ & $ i \in \{OEGMA\}$  \\
\end{tabular}
\end{center}
\end{table}

Specifically,
$\beta_{S1} $ is for categorical distribution over $\mathcal{S}_y$ when $z_k = S1$. 
$\beta_{S2} $ is for categorical distribution over $\mathcal{S}_y  \setminus \{SPMA\}$ when $z_k = S2$, because we assume the SPMA monomer appears only in $S1$ segments. 
$\beta_{S3} $ is for categorical distribution over $\mathcal{S}_y  \setminus \{OEGMA, SPMA\}$ if $z_k = S3, \tau_k \neq\iota_k$ or $z_k = S3, \iota_k = 0$. It is used when the current position $k$ is within an $S3$ segment but no OEGMA is emitted, either because $\tau_k$ hasn't reached the designated OEGMA location $\iota_k$ or because the current $S3$ segment is made up of purely MMA and EHMA.
$\beta_{S3'}$ is for categorical distribution over $\{OEGMA\}$ if $z_k = S3, \tau_k=\iota_k$. It degenerates to a deterministic value. It is used when the duration variable $\tau_k$ decreases to the designated OEGMA location $\iota_k$, indicating that the only OEGMA should be emitted in the current $S3$ segment. Thus we have:
\begin{equation}
\begin{aligned}
\label{eq:emission}
& \ln 
[p(y_k | z_k, \tau_k, \iota_k, \beta)] 
\\
=
 & \sum_{j \in \mathcal{S}_y} \sum_{\text{m} \in \{S1, S2\} } \Big[ \ln(\beta_{\text{m}, j}) 
  \mathbbm{1}\{ z_k = \text{m}, y_k = j\} \Big] +
    \sum_{j \in \mathcal{S}_y} \sum_{\text{m} \in \{S3'\} } \Big[ \ln(\beta_{\text{m}, j}) 
   \mathbbm{1}\{z_k = m, \tau_k=\iota_k, y_k = j\} \Big] + 
   \\
 &     \sum_{j \in \mathcal{S}_y} \sum_{\text{m} \in \{S3\} } \Big[ \ln(\beta_{\text{m}, j})
   \mathbbm{1}\{z_k = m, \tau_k \neq \iota_k  \text{ or }  z_k = m, \iota_k = 0, y_k = j\} \Big]. 
\end{aligned}
\end{equation}

This probabilistic model incorporates the aforementioned constraints on RHP systems, and allows for explicit modeling of the hidden states' duration. The complete likelihood function for one RHP chain is:
\begin{equation}
\begin{aligned}
\label{eq:complete likelihood}
 & p(y_{1:K}, z_{1:K}, \tau_{1:K}, \iota_{1:K} | \pi,  \alpha, \beta, \rho)   
 \\
 = & p(z_1 | \pi) p(\tau_1, \iota_1 |z_1, \rho) 
  \prod_{k=2}^K  [p( z_k, \tau_k, \iota_k |  z_{k-1}, \tau_{k-1}, \iota_{k-1},   \alpha, \rho) ] 
 \prod_{k=1}^K[p(y_k | z_k, \tau_k, \iota_k, \beta)] 
 \\
 = & \exp \{ {\eta}({\theta})^\top \cdot {\phi}(\bm{z, \tau, \iota, y})  \}. 
\end{aligned}
\end{equation}

In the last line of Equation \ref{eq:complete likelihood}, we plug in Equations \ref{eq:initial} to \ref{eq:emission} and 
simplify Equation \ref{eq:complete likelihood} further to an exponential family.
Here, ${\theta} = (\pi, \alpha, \beta, \rho)$. ${\eta}({\theta}) = \ln \theta$ is the canonical parameter vector and $\ln$ is applied element-wise.
We denote $\bm{y} = y_{1:K} = (y_1,...,y_K)$, with $\bm{z, \tau, \iota}$ similarly defined. $ {\phi}(\bm{z, \tau, \iota, y})$ is the vector of complete-data sufficient statistics. 
All sequences are i.i.d. given the parameter ${\theta}$,
but for clarity we have suppressed indices for i.i.d. replicas of RHP sequences so far. 
Denote $\bm{y}^{(s)} = y_{1:K}^{(s)} = (y_1^{(s)},...,y_K^{(s)})$ as observations of the $s$-th sequence, $s = 1,...,N$.
Denote $\bm{y}^{(\mathcal{I})} = \{ \bm{y}^{(s)} \}_{s \in \mathcal{I}}$ as all RHP sequences in the index set $\mathcal{I} \subseteq \{1,...,N\}$, i.e., $\bm{y}^{(1:N)} = \{ \bm{y}^{(s)} \}_{s=1}^N$. $\bm{z}^{(s)}, \bm{\tau}^{(s)}, \bm{\iota}^{(s)}$ and $\phi^{(s)} = \phi( \bm{z}^{(s)}, \bm{\tau}^{(s)}, \bm{\iota}^{(s)}, \bm{y}^{(s)} )$ are defined similarly. 

\section{Stochastic Variational Methods for the GHSMM-RHP}
\label{sec:Method}

Variational methods have been used for approximate inference under different scenarios when exact solutions are intractable or too expensive to achieve \citep{bishop2006pattern, wainwright2008graphical}.
To develop more efficient algorithms in modern large-scale learning problems, stochastic optimizations are introduced to bypass the need to iterate through the entire datasets in each iteration \citep{robbins1951stochastic}.
Combining variational methods and stochastic optimizations helps alleviate the computational complexity of estimations in the GHSMM-RHP model.

In particular, 
in the Bayesian setting where parameters $\theta$ are viewed as random variables, we coupled stochastic natural gradient descent \citep{amari1998natural} with variational Bayes following the work of \cite{hoffman2013stochastic}.
In the frequentist setting where parameters $\theta$ are viewed as fixed but unknown, we coupled the stochastic Frank-Wolfe algorithm \citep{reddi2016stochastic} with variational expectation-maximization.

\subsection{Stochastic Variational Bayesian}

In this section, we describe the stochastic variational Bayesian (SVB) algorithms for our GHSMM-RHP model. For the complete likelihood function $p( \bm{z}^{(1:N)}, \bm{\tau}^{(1:N)}, \bm{\iota}^{(1:N)}, \bm{y}^{(1:N)}  | {\theta}) = \exp \{  {\eta}({\theta})^\top \sum_{s=1}^N \phi^{(s)}  \} $, we consider the conditionally conjugate exponential family \citep{bernardo2009bayesian} with priors taking the form of $p({\theta} | \widetilde{\lambda}_{\theta} ) = \exp \{ {\widetilde{\lambda}}_{\theta}^\top {\eta}({\theta}) - B(\widetilde{\lambda}_{\theta}) \}$.
Here, prior parameters $\widetilde{\lambda}_{\theta} = ( {\widetilde{\lambda}}_{\pi}, {\widetilde{\lambda}}_{\alpha},  {\widetilde{\lambda}}_{\beta}, {\widetilde{\lambda}}_{{\rho}})$ are for ${\theta} = (\pi, \alpha, \beta, \rho)$, respectively.
These add to the full model with:
\begin{equation}
\begin{aligned}
\label{eq:global latent prior}
{\pi} | {\widetilde{\lambda}}_{\pi} \sim \text{Dir}({\widetilde{\lambda}}_{\pi} + \bm{1}), 
 \\
\alpha_i | {\widetilde{\lambda}}_{\alpha_i} \sim \text{Dir}({\widetilde{\lambda}}_{\alpha_i} + \bm{1}), \forall i \in \{S1, S2, S3\},  
 \\ 
{\rho}^i | {\widetilde{\lambda}}_{{\rho}^i} \sim \text{Dir}({\widetilde{\lambda}}_{{\rho}^i} + \bm{1}) , \forall i \in  \{S1, S2, S3\},  
 \\ 
\beta_{\text{m}} | {\widetilde{\lambda}}_{\beta_{\text{m}}} \sim \text{Dir}({\widetilde{\lambda}}_{\beta_{\text{m}}} + \bm{1}), \forall \text{m} \in  \{S1, S2, S3, S3'\}. 
 \\ 
\end{aligned}
\end{equation}
$S3$ and $S3'$ refer to the same segment as described in Table \ref{tab:emis_prob_para}.
The complete conditional distribution of local latent variables can be written from Equation \ref{eq:complete likelihood} as:
\begin{equation}
\begin{aligned}
\label{eq:local latent posterior}
  p(\bm{z}^{(1:N)}, \bm{\tau}^{(1:N)}, \bm{\iota}^{(1:N)} | \theta, \bm{y}^{(1:N)}) 
= & \prod_{s=1}^N p(\bm{z}^{(s)}, \bm{\tau}^{(s)}, \bm{\iota}^{(s)} | \theta, \bm{y}^{(s)}) 
 \\
= & \prod_{s=1}^N \exp \{ {\eta}({\theta})^\top  \phi^{(s)} - A_{y^{(s)}}(\eta(\theta)) \}. 
 \\
\end{aligned}
\end{equation}
Here, the cumulant function $A_{y^{(s)}}(\eta(\theta)) = \ln \sum_{\bm{z}^{(s)}, \bm{\tau}^{(s)}, \bm{\iota}^{(s)}} \exp \{ {\eta}({\theta})^\top  \phi^{(s)} \}$. 
Define its conjugate function as $A^*_{y^{(s)}}(\mu) = \max_{\eta} ( \mu^\top \eta - A_{y^{(s)}}(\eta))$. Note the analogue term for all N sequences is $A_{y^{(1:N)}}(\eta(\theta)) = \sum_{s=1}^N A_{y^{(s)}}(\eta(\theta))$.

We then specify a family of variational distributions $q(\bm{z}^{(1:N)}, \bm{\tau}^{(1:N)}, \bm{\iota}^{(1:N)}, \theta | \psi, \lambda_\theta) \in \mathcal{Q}$, aiming to approximate the true posterior distribution over all latent variables in terms of KL divergence. We make the mean-field assumption in the variational family $\mathcal{Q}$: independence between $\theta$ and $(\bm{z}^{(1:N)}, \bm{\tau}^{(1:N)}, \bm{\iota}^{(1:N)})$ while structures are kept within $(z_{1:K}^{(s)}, \tau_{1:K}^{(s)}, \iota_{1:K}^{(s)})$, namely $q(\bm{z}^{(1:N)}, \bm{\tau}^{(1:N)}, \bm{\iota}^{(1:N)}, \theta) = q(\theta) \prod_{s=1}^N q(\bm{z}^{(s)}, \bm{\tau}^{(s)}, \bm{\iota}^{(s)})$. It can be argued that the optimal variational distribution takes the same forms of exponential families as in Equations \ref{eq:global latent prior} and \ref{eq:local latent posterior} \citep{bishop2006pattern}. That is, 
$q(\theta | \lambda_{\theta}) = \exp \{ \lambda_{\theta}^\top \eta(\theta) - B({\lambda}_{\theta}) \}$, $q( \bm{z}^{(s)}, \bm{\tau}^{(s)}, \bm{\iota}^{(s)} | \psi) =  \exp \{ \psi^\top \phi^{(s)} - A_{y^{(s)}}(\psi) \}, \forall s \in 1,...,N$. Here, different RHP sequences share the same local variational parameters $\psi$. The sequence specific information is decoupled into sufficient statistics $\phi^{(s)}$. Using $ \mathbb{E}_{\lambda_\theta}, \mathbb{E}_{\psi}$ to denote the corresponding expectations, the evidence lower bound function (ELBO) is:
\begin{equation}
\begin{aligned}
\label{eq:SVB ELBO}
 \ln p(\bm{y}^{(1:N)})  \geq &  ELBO_{SVB} 
 \\
:= & \ln p(\bm{y}^{(1:N)}) - \text{KL} (q(\bm{z}^{(1:N)}, \bm{\tau}^{(1:N)}, \bm{\iota}^{(1:N)}, \theta | \psi, \lambda_\theta) 
  || p(\bm{z}^{(1:N)}, \bm{\tau}^{(1:N)}, \bm{\iota}^{(1:N)}, \theta | \bm{y}^{(1:N)}))) 
 \\
 = &   \Big \langle \sum_{s=1}^N \mathbb{E}_{\psi} {\phi}^{(s)} , \mathbb{E}_{\lambda_\theta} \eta(\theta) \Big \rangle - A_{y^{(1:N)}}^*( \sum_{s=1}^N \mathbb{E}_{\psi} {\phi}^{(s)} )  - \text{KL}( q_{\lambda_\theta}|| p_{\widetilde{\lambda}_{\theta}} )  - 
 \sum_{s=1}^N \mathbb{E}_{\lambda_\theta} A(\eta(\theta)), 
   \\
\end{aligned}
\end{equation}
where $\langle \cdot, \cdot \rangle$ denotes inner product.
The last line is based on specific forms of variational distributions.
Note here the cumulant function $A(\eta({\theta})) = \ln \sum_{\bm{y}^{(s)}, \bm{z}^{(s)}, \bm{\tau}^{(s)}, \bm{\iota}^{(s)}} \exp \{ {\eta}({\theta})^\top  \phi^{(s)} \} = 0$ with a probability of one, if $\theta$ follows Dirichlet distributions.
Maximizing the ELBO function is equivalent to finding the best approximation in $\mathcal{Q}$ to the true posterior distribution. One way to do this is through stochastic natural gradient descent \citep{hoffman2013stochastic} while iteratively updating the local variational distribution $q_{\psi^{(t)}} := q(\bm{z}^{(1:N)}, \bm{\tau}^{(1:N)}, \bm{\iota}^{(1:N)} | \psi^{(t)})$ and the global variational density $q_{\lambda_\theta^{(t)}} := q(\theta | \lambda_{\theta}^{(t)})$. Here $(t)$ is the number of iterations.

E-step: solving $\max _{\psi} ELBO_{SVB} \big(  q_\psi, q_{\lambda_\theta^{(t)}}  \big) = \max _{\psi} \big(  \langle  \sum_{s=1}^N \mathbb{E}_{\psi} {\phi}^{(s)} , \mathbb{E}_{\lambda_\theta^{(t)}} \eta(\theta)  \rangle -  \linebreak A_{y^{(1:N)}}^*( \sum_{s=1}^N \mathbb{E}_{\psi} {\phi}^{(s)} )  \big) $. Convexity in $ \mathbb{E}_{\psi} {\phi}^{(s)}$ yields the update $\psi^{(t)} = \mathbb{E}_{\lambda_\theta^{(t)}} \eta(\theta)$.

M-step: solving $\max _{\lambda_\theta} ELBO_{SVB} \big(  q_{\psi^{(t)}} , q_{\lambda_\theta}   \big) = \max _{\lambda_\theta} \big( \langle \sum_{s=1}^N  \mathbb{E}_{\psi^{(t)}} {\phi}^{(s)} , \mathbb{E}_{\lambda_\theta} \eta(\theta) \rangle  - \linebreak  \text{KL}( q_{\lambda_\theta}|| p_{\widetilde{\lambda}_{\theta}} )  \big)$. 
Using stochastic natural gradient descent yields the update:
\begin{equation}
\begin{aligned}
\label{eq:SVB M-step}
\lambda_\theta^{(t+1)} = (1-\gamma^{(t)}) \lambda_\theta^{(t)} + \gamma^{(t)} ( \frac{N}{|\mathcal{I}|} \sum_{s \in \mathcal{I}} \mathbb{E}_{\psi^{(t)}} {\phi}^{(s)} + \widetilde{\lambda}_\theta),
\end{aligned}
\end{equation}
where $\mathcal{I}$ is the randomly chosen mini-batch and $\gamma^{(t)}$ is the step size at the $t$-th iteration.
To calculate $ \mathbb{E}_{\psi^{(t)}} {\phi}^{(s)} $, 
we implemented a message-passing recursion
which is a special version of the junction tree algorithm \citep{lauritzen1988local}. More details on 
update formulas can be found in Supplementary Materials section B.

\subsection{Stochastic Variational Expectation Maximization}

Under the frequentist setting,
the complete likelihood $p( \bm{z}^{(1:N)}, \bm{\tau}^{(1:N)}, \bm{\iota}^{(1:N)}, \bm{y}^{(1:N)}  | {\theta})$ in Equation \ref{eq:complete likelihood},
complete conditional distributions $p( \bm{z}^{(1:N)}, \bm{\tau}^{(1:N)}, \bm{\iota}^{(1:N)} | \bm{y}^{(1:N)}, {\theta})$ in Equation \ref{eq:local latent posterior}, and local variational distributions $q( \bm{z}^{(s)}, \bm{\tau}^{(s)}, \bm{\iota}^{(s)} | \psi)$
remain the same as those in SVB.
Thus, stochastic variational expectation-maximization (SVEM) for the GHSMM-RHP model carries similar ELBO and parameter updates to the Bayesian approach:
\begin{equation}
\begin{aligned}
\label{eq:SVEM ELBO}
\ln p(\bm{y}^{(1:N)} | \theta)  \geq  & ELBO_{SEM} 
 \\
:= & \ln p(\bm{y}^{(1:N)} | \theta) - \text{KL} (q(\bm{z}^{(1:N)}, \bm{\tau}^{(1:N)}, \bm{\iota}^{(1:N)} | \psi) 
  || p(\bm{z}^{(1:N)}, \bm{\tau}^{(1:N)}, \bm{\iota}^{(1:N)}| \theta, \bm{y}^{(s)}))) 
 \\
 = &  \Big \langle \sum_{s=1}^N \mathbb{E}_{\psi} {\phi}^{(s)} , \eta(\theta) \Big \rangle - A_{y^{(1:N)}}^*( \sum_{s=1}^N \mathbb{E}_{\psi} {\phi}^{(s)} ) - \sum_{s=1}^N A(\eta(\theta)).  
  \\
\end{aligned}
\end{equation}

E-step: solving $\max_{\psi} ELBO_{SEM} \big(  q_\psi, \theta^{(t)}  \big) = \max _{\psi} \big( \langle  \sum_{s=1}^N \mathbb{E}_{\psi} {\phi}^{(s)} ,  \eta(\theta^{(t)})  \rangle - \linebreak A_{y^{(1:N)}}^*( \sum_{s=1}^N \mathbb{E}_{\psi} {\phi}^{(s)} ) \big) $, yielding the update $\psi^{(t)} = \eta(\theta^{(t)})$.

M-step: solving $\max_{\theta} ELBO_{SEM} \big(  q_{\psi^{(t)}}, \theta   \big) = \max _{\theta} \big(  \langle \sum_{s=1}^N  \mathbb{E}_{\psi^{(t)}} {\phi}^{(s)} ,  \eta(\theta) \rangle - \sum_{s=1}^N A(\eta(\theta)) \big) $. Using the stochastic Frank-Wolfe algorithm yields the update:
\begin{equation}
\begin{aligned}
\label{eq:SVEM M-step}
\theta^{(t+1)} &= (1-\gamma^{(t)}) \theta^{(t)} + \gamma^{(t)} e_j   ,
 \\
j &=  \arg\max  \frac{ \frac{N}{|\mathcal{I}|} \sum_{s \in \mathcal{I}} (\mathbb{E}_{\psi^{(t)}} {\phi}^{(s)} - \mathbb{E}_{y, z, \tau, \iota | \eta(\theta^{(t)})}\phi^{(s)}) }{\theta^{(t)}}   ,
 \\
\end{aligned}
\end{equation}
where $e_j$ is the basis vector with 1 at the j-th element and 0 otherwise, $\mathcal{I}$ is the randomly chosen mini-batch, and $\gamma^{(t)}$ is the step size at the $t$-th iteration.

There are some notable differences with the SVB algorithm.
In SVB,
the cumulant function $A(\eta({\theta})) = \ln \sum_{\bm{y}^{(s)}, \bm{z}^{(s)}, \bm{\tau}^{(s)}, \bm{\iota}^{(s)}} \exp \{ {\eta}({\theta})^\top  \phi^{(s)} \} = 0$. 
The M-step involves evaluating $\mathbb{E}_{\psi^{(t)}} {\phi}^{(s)}$ with respect to the local variational distribution 
$q(\bm{z}^{(s)}, \bm{\tau}^{(s)}, \bm{\iota}^{(s)} | \psi^{(t)})$.
In SVEM,
the cumulant function $A(\eta({\theta})) \neq 0$ for parameters $\theta$ that consist of sub-normalized probability vectors (e.g., the sum of elements in updated $\pi^{(t)}$ is not one). 
Besides the message-passing recursion used in SVB, 
the M-step in SVEM requires a new collecting-distributing pass
to calculate the unconditional expectation $\mathbb{E}_{y, z, \tau, \iota | \eta(\theta^{(t)})}\phi^{(s)}$ regarding the complete likelihood $p( \bm{z}^{(s)}, \bm{\tau}^{(s)}, \bm{\iota}^{(s)}, \bm{y}^{(s)}  | {\theta^{(t)}})$.
More details 
on message-passing 
can be found in Supplementary Materials section A.
Additionally, SVEM involves constrained optimizations on probability simplexes in order to guarantee the interpretability of parameter estimates (i.e., the sum of elements in $\pi^{(t)}$ remains one).
In SVB, 
parameter constraints are automatically guaranteed after every update.

\section{Results}
\label{sec:Result}

We implemented SVB and SVEM, and compared their performance using a comprehensive simulation study and a real data study from \cite{jiang2020single}.

\subsection{Simulation Study}

\subsubsection{Setup}
\label{sec:simulation setup}

\textit{\textbf{Data Generation}}
This study contains 20 simulated heteropolymers (SHP) datasets labeled SHP1, ..., SHP20.
They were generated following Equation \ref{eq:complete likelihood}.
The study combines ten sets of predetermined parameters $(\pi, \alpha, \beta)$ and two different values of $\underline{D}$, the shortest lengths allowed for $S3$ segments.
Table \ref{tab:SHP} illustrates parameters for SHP1 and SHP2.
In the simulation,
each set of $(\pi, \alpha, \beta)$ was combined with two values of $\underline{D}$ to generate a pair of ``twin'' SHP datasets, e.g., SHP1 and SHP2 form a ``twin'' pair, SHP3 and SHP4 form a ``twin'' pair, et cetera. That is, each ``Twin" pair of SHPs share the same initial probabilities $\pi$, transition probabilities $\alpha$, and emission probabilities $\beta$, but differ in the duration probabilities $\rho$ since it relies on $\underline{D}$. Thus, twin SHPs are more similar to each other than to the rest of data sets. 
This design ensures that the simulated SHP systems have a wide range of heterogeneity, with some being more similar and others being very different.

\begin{table}
\caption{Simulation for one pair of ``twin'' SHPs.}
\label{tab:SHP}
\begin{center}
\begin{tabular}{| c | c | c | c |}
\hline
Simulated data   & Shared parameters & $\underline{D}$ & Duration parameters \\
\hline
SHP1 &   \multirow{2}{*}{$(\pi, \alpha, \beta) \sim \text{Dir}(\bm{1}) $}   &  7 & $ \rho  \sim \text{Dir}_{\underline{D}}(\bm{1})$ \\
SHP2 &                                                                                                   &  9 & $ \rho  \sim \text{Dir}_{\underline{D}}(\bm{1})$   \\
\hline
\end{tabular}
\end{center}
\end{table}

In more detail, elements in every set of $\theta = (\pi, \alpha, \beta, \rho)$ were sampled from the $\text{Dir}(\bm{1})$ distribution. 
Recall that the supports for parameters $\rho = (\rho^{S1}, \rho^{S2}, \rho^{S3})$,
namely $\mathcal{S}^{S1}_{\tau,\iota}, \mathcal{S}^{S2}_{\tau,\iota}$ or $\mathcal{S}^{S3}_{\tau,\iota}$,
rely on $\underline{D}$ and $\overline{D}$ (hence the notation $\text{Dir}_{\underline{D}}(\bm{1})$ in Table \ref{tab:SHP}).
$\underline{D}$ was set to 7 or 9,
and $\overline{D}$ was fixed to 25.
These values are based on the current knowledge of RHP systems.
For each parameter setup ($\underline{D}$ and $\theta$),
we generated one SHP dataset comprised of 5000 chains, each chain at a length of 130 monomers.
In each SHP dataset, 500 chains were held out and used as the test set, while the other 4500 chains were used as the training set.

\textit{\textbf{Training Settings}}
For each of the 20 simulated datasets, both SVB and SVEM algorithms were applied with various hyper-parameters.
Specifically, 
we used 10 random initializations in each algorithm. For every initialization, we tried 15 learning rates: in SVB 
the step size was $\gamma^{(t)} = \frac{1}{(t+\kappa_1-1)^{\kappa_2}}$ where
$\kappa_1 \in \{10^3, 10^5, 10^6, 10^7, 10^9\}$,
$\kappa_2 \in \{1, 0.9, 0.7\}$;
in SVEM 
the step size was $\gamma^{(t)} = \frac{2}{(t+2+\kappa_1-1 )^{\kappa_2}}$ where
$\kappa_1 \in \{10^0, 10^1, 10^2, 10^4, 10^6\}$,
$\kappa_2 \in \{1, 0.9, 0.7\}$.
They become the vanilla stochastic (conditional) gradient method when the ``decay'' $\kappa_1$ equals $10^0$ and the ``rate'' $\kappa_2$ equals 1. 
During training we also explored varying values of batch size $| \mathcal{I} | \in \{12,48,120\}$ (in Equations \ref{eq:SVB M-step} and \ref{eq:SVEM M-step}) and shortest $S3$ segment length $\underline{D} \in \{5,7,10,15\}$. 
However, we found that neither algorithm was sensitive to these two hyper-parameters, and we do not report them here for brevity. 
The choice of $| \mathcal{I} |$ and $\underline{D}$ were fixed to 48 and 7, respectively.

In summary, we had 150 training settings. 
Under each of the 10 initializations, we tried 15 learning rates and reported the one that gave the highest value in objective functions ELBO.

In the following subsections, we empirically studied the two stochastic variational methods and showcased the computational instability rooted in SVB. 
To our best knowledge, there has been little effort to compare their performance or theoretically characterize their convergence behavior in the literature.
Our comparisons between SVB and SVEM in the RHP applications may shed light on an understanding of Bayesian versus frequentist stochastic variational methods.

\subsubsection{Performance Evaluation}
\label{sec:Result:se}

\begin{figure}
\centering
     \begin{subfigure}[b]{0.49\textwidth}
         \centering
         \includegraphics[width=\textwidth]{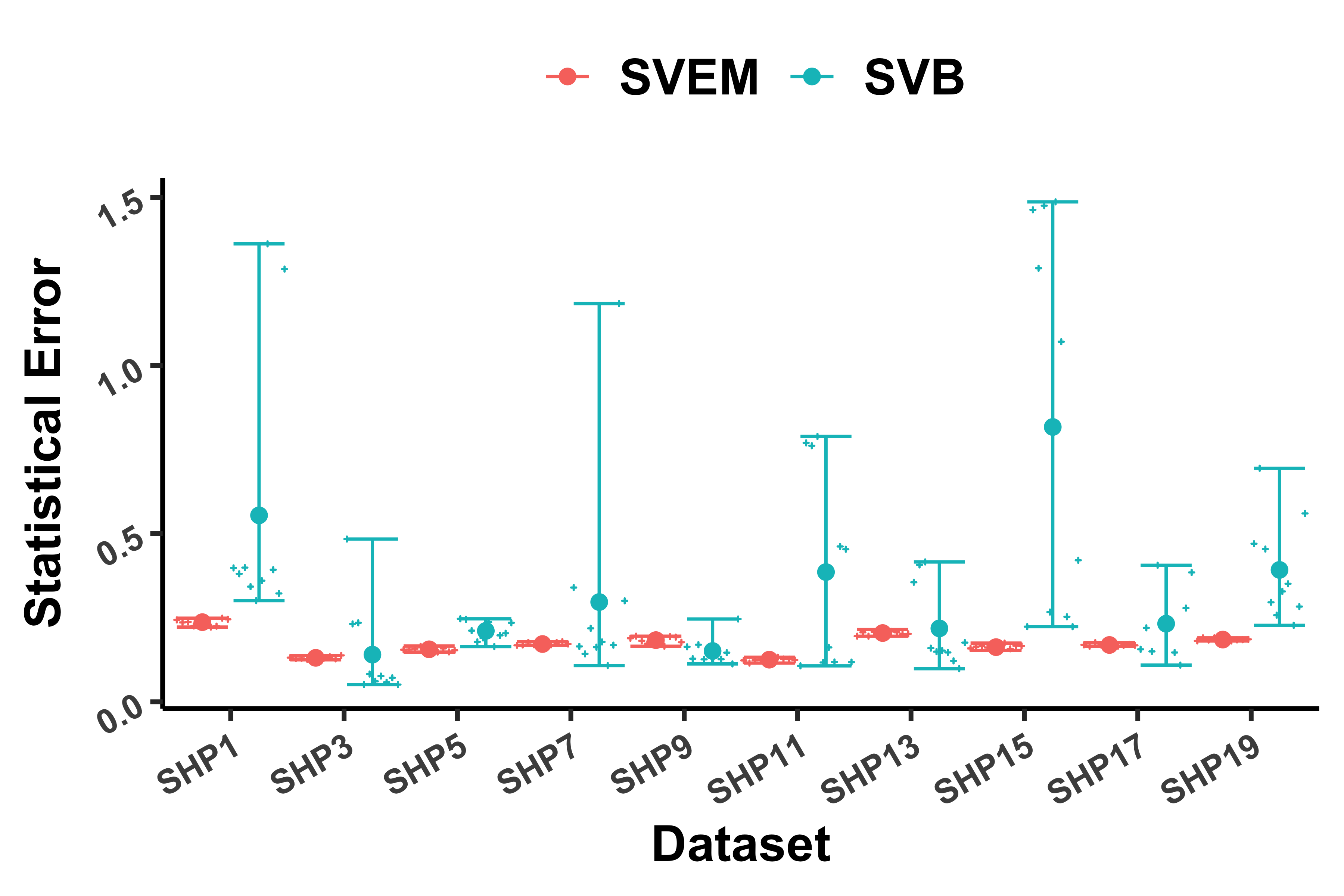}
     \end{subfigure}
\caption{Statistical error of two point estimates: SVB in blue and SVEM in red. It plots $|| \hat{\theta} - \theta^{\ast} ||_2$ (y-axis) under different SHPs (x-axis). Only one out of every pair of ``twin'' SHPs is presented. 
The estimates were chosen after 1000 iterations for both algorithms, namely $\hat{\theta} = \hat{\theta}^{(1000)}$. $\theta^{\ast}$ is the true parameter simulated in Section \ref{sec:simulation setup},
e.g., $\theta^{\ast}_{Shp1}$ for SHP1, $\theta^{\ast}_{Shp3}$ for SHP3.
Every individual bar contains 10 values of statistical error, 
based on $\hat{\theta}$ trained from 10 initialization.
The mean and range of the 10 values are shown.}
\label{fig:se}
\end{figure}

\textit{\textbf{Point Estimation.}} 
Through the GHSMM-RHP model, we obtained two point estimates for $\theta$.
We implemented SVB and used the posterior mean $ \mathbb{E}_{ \lambda_\theta^{(t)} } \theta$ as estimates under the Bayesian framework.
We implemented SVEM and used approximated maximum likelihood estimates $\theta^{(t)}$ under the frequentist framework.
Figure \ref{fig:se} shows the statistical error of the two estimates in different SHPs.
For simplicity only one out of every pair of ``twin'' SHPs is presented, 
because we observed that
``twin'' data sets, such as SHP1 and SHP2, had almost identical results. This indicates that the variational objectives of our model (i.e., Equations \ref{eq:SVB ELBO} and \ref{eq:SVEM ELBO}) are not sensitive to the parameter $\underline{D}$.\footnote{This relates to the formulation of objective function itself (for more discussion, see for example \citep{giordano2018covariances}; \citep[Chapter~7]{wainwright2008graphical}), and is beyond the scope of current work focusing on algorithmic issues.}

Figure \ref{fig:se} shows that SVEM consistently produces low statistical error that is comparable to the best in SVB.
However, the SVB algorithm appears more sensitive to the local optima issue and may require ``better'' initializations (e.g., to be closer to the truth) than the SVEM,
even though the global optimizer of variational objectives present arguably good statistical properties \citep{wang2019frequentist}.
That is, to obtain a near-optimal solution in SVB, 
more tuning efforts may be needed, such as trying more random initializations or using a proper pilot algorithm. Additionally, we observed that it took fewer iterations for SVB to converge in this study.
Note that the running time per iteration is of the same order for both methods, because roughly equal numbers of message passes are required in the E-step.

\begin{figure}
\centering
     \begin{subfigure}[b]{0.49\textwidth}
         \centering
         \includegraphics[width=\textwidth]{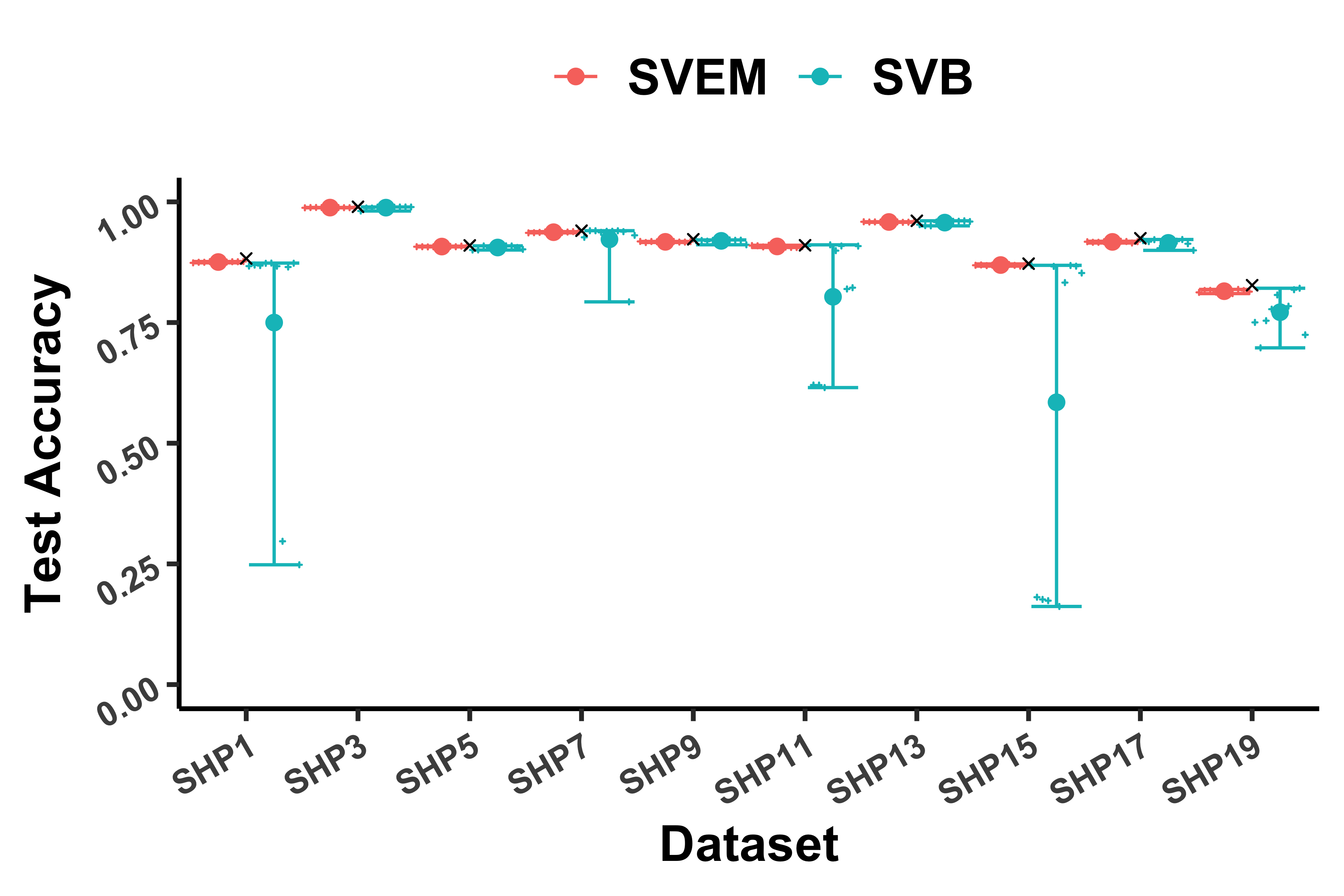}
         \caption{\textbf{Overall accuracy for $\bm{S1, S2, S3}$.}}
         \label{pred_S123.png}
     \end{subfigure}
     \begin{subfigure}[b]{0.49\textwidth}
         \centering
         \includegraphics[width=\textwidth]{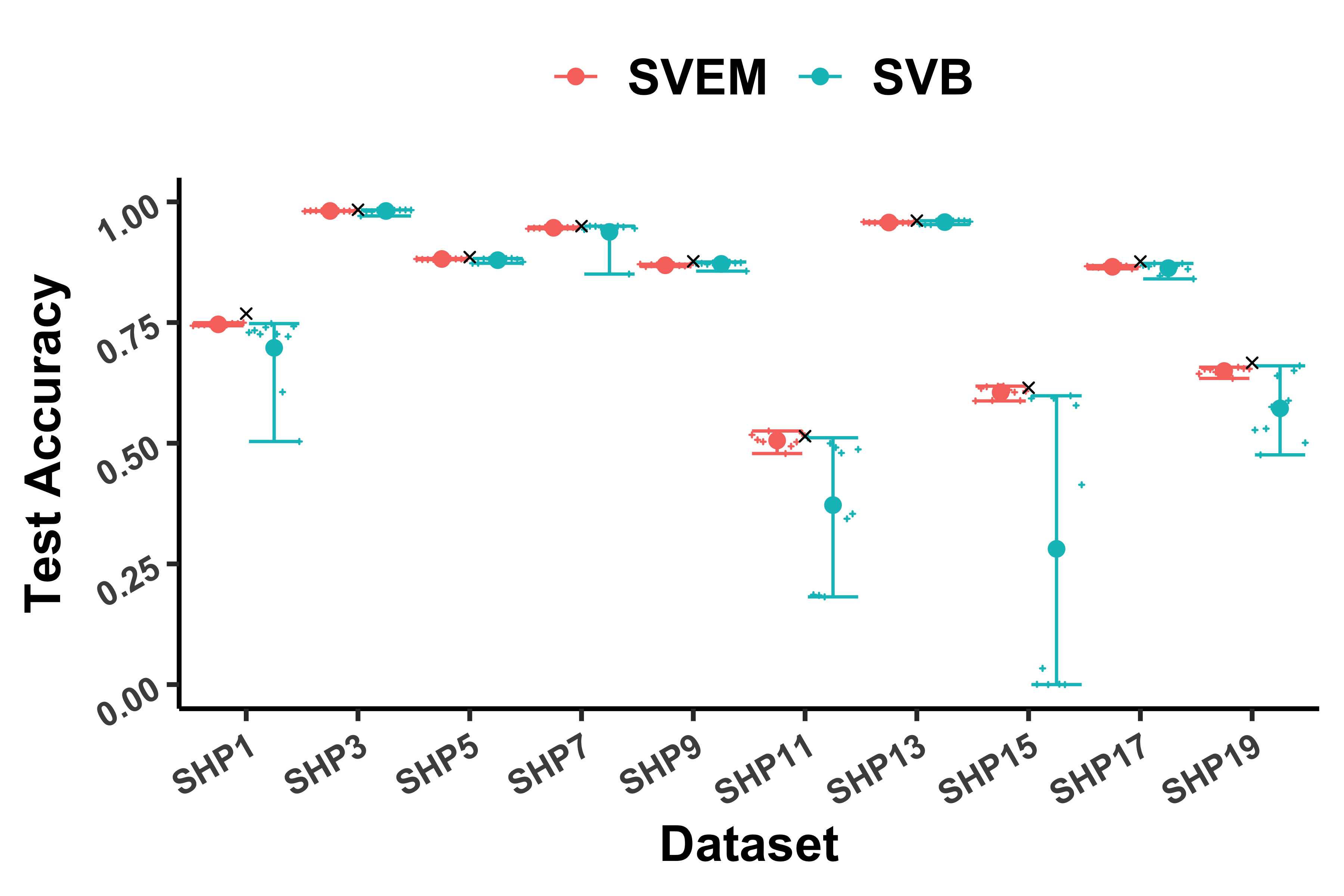}
         \caption{\textbf{Jaccard index for $\bm{S3}$.}}
         \label{pred_S3.png}
     \end{subfigure}
\caption{Segment prediction using two point estimates: SVB in blue and SVEM in red. 
The prediction was done on each SHP (labeled on x-axis, training and test sets were from the same SHP dataset), but only one out of every pair of ``twin'' SHPs is presented. 
Every individual bar contains 10 values of test accuracy, based on estimates trained from 10 initializations.
The mean and range of the 10 values are shown. 
In panel (a), the prediction accuracy is for all segments $S1, S2$, and $S3$ based on the 0-1 loss function. It simply calculates the proportion of monomers correctly predicted when compared to the true labels. Panel (b) presents the Jaccard index concerning solely the accuracy of hydrophobic segment $S3$.
It calculates the ratio of two numbers: the correctly predicted $S3$ monomers, the union of ground truth $S3$ and predicted $S3$ monomers.
Black cross dots marks the test accuracy under true parameters of each SHP, namely $\theta^{\ast}_{Shp1}$, ..., $\theta^{\ast}_{Shp19}$.
}
\label{fig:S123}
\end{figure}

\textit{\textbf{Segment Prediction.}}
We followed the Viterbi algorithm \citep{viterbi1967error} and predicted the hidden segment states on each test set of the 20 SHPs, using parameter estimates from SVB and SVEM.
Figure \ref{fig:S123} summarizes the findings.
As in Figure \ref{fig:se}, only one out of every pair of ``twin'' SHPs is presented because ``twin'' data sets have similar results.

Panel (a) measures the overall prediction accuracy for segment states $S1, S2$, and $S3$. Panel (b) measures the accuracy for $S3$ segments alone. 
Compared to $S1$ and $S2$, $S3$ appears less frequently and is more functionally interesting because it has the potential to form a transportation tunnel. 
Figure \ref{fig:S123} shows that SVB and SVEM produced comparable accuracy if trained with proper initializations.
Note that SHP11 and SHP15 present lower Jaccard index for $S3$ segments (shown in panel (b)) when compared to the overall accuracy (shown in panel (a)).
We observed that there are fewer hydrophobic segments $S3$ in SHP11 and SHP15 than in the other SHPs.
So, even if $S3$ predictions are relatively poor in SHP11 and SHP15, the overall accuracy illustrated in panel (a) is not affected much due to the dominance of $S1$ and $S2$ in the chains.

Figure \ref{fig:S123} again shows that predictions (using the Viterbi algorithm) by SVB estimates remain less robust to random initializations, while SVEM produces consistent better predictions assessed by overall accuracy and Jaccard index.
However in some datasets (e.g., SHP3), the segment prediction based on SVB estimates remains stable (shown in Figure \ref{fig:S123}) despite SVB's lack of robustness in terms of statistical error (shown in Figure \ref{fig:se}).
The black cross dots in Figure \ref{fig:S123} are the prediction accuracy using the true parameters for each SHP, i.e., $\theta^{\ast}_{Shp1}$, ..., $\theta^{\ast}_{Shp19}$.
As can be seen, the accuracy based on SVB and SVEM estimates is comparable to the accuracy based on true parameters.

\subsubsection{Applications} 
\label{sec:lpp}

Designing bio-functional polymers relies on understanding the heterogeneity of RHP chains under various overall monomer distributions.
Through GHSMM-RHP, we can statistically quantify the relationships among different RHP datasets
using posterior predictive probability in the Bayesian scheme: $p(\bm{y}^{(new)} | \bm{y}^{(1:N)}) =  \int p(\bm{y}^{(new)} | \theta) p(\theta | \bm{y}^{(1:N)})  d \theta $, which is the average likelihood of the new sequences $\bm{y}^{(new)}$ based on distributions of $\theta$ learned from the observed data.
Greater similarity indicates that new sequences may share more similar chemical properties with the training sequences.
Several lines of empirical research have shown that the predictive inference based on variational Bayes is similar to the fully Bayes predictive inference based on Markov chain Monte Carlo (MCMC) \citep{blei2006variational, braun2010variational}.
\cite{nott2012regression} also used variational approximations in the posterior predictive probability for a cross-validation based model selection.

There are several ways to approximate the intractable posterior predictive probability. 
One popular method is to
replace $p(\theta | \bm{y}^{(1:N)})$ with the variational posterior $q(\theta | \lambda_\theta^{(t)} )$ and construct a Monte Carlo approximation $p(\bm{y}^{(new)} | \bm{y}^{(1:N)}) \approx  \int q(\theta | \lambda_\theta^{(t)} ) p(\bm{y}^{(new)} | \theta) d \theta $. 
The time complexity is high, however, 
since it requires the same number of message-passings as the Monte Carlo samples. 
\cite{beal2003variational} proposed a simple lower bound whose computation
requests only one round of message-passing as shown in Supplementary Materials section A. 
However, it is based on a bound of the approximation to the posterior predictive probability, which may raise concern in some situations.
The third method is to concentrate the local variational distribution on its posterior mean, a point mass $ \mathbb{E}_{ \lambda_\theta^{(t)} } \theta$, such that $p(\bm{y}^{(new)} | \bm{y}^{(1:N)}) \approx  p(\bm{y}^{(new)} | \mathbb{E}_{ \lambda_\theta^{(t)} } \theta) $. 
The computation requires only one message-passing, the same as in the second method.

Our simulations showed the above three methods gave similar numerical values to the posterior predictive probability. 
We chose the third method $p(\bm{y}^{(new)} | \bm{y}^{(1:N)}) \approx  p(\bm{y}^{(new)} | \mathbb{E}_{ \lambda_\theta^{(t)} } \theta) $ as it avoids the sample based time complexity in the first method, 
and serves as a reliable approximation.
$ \mathbb{E}_{ \lambda_\theta^{(t)} } \theta$ is referred to as variational Bayes estimate in \cite{wang2019frequentist}, and is closely related to the SVEM estimates $\theta^{(t)}$ discussed in Section \ref{sec:Result:se}. 
Thus under the frequentist framework, the approximated maximum likelihood $p(\bm{y}^{(new)} | \theta^{(t)}) $ offers an analogous measure, 
and draws an easier comparison of likelihood between the two algorithms. That is, $\ln p(\bm{y}^{(new)} | \mathbb{E}_{ \lambda_\theta^{(t)} } \theta)$ in SVB and $\ln p(\bm{y}^{(new)} | \theta^{(t)}) $ in SVEM, which are both referred to as log predictive probability in the following discussions.

\begin{figure}
\centering
     \begin{subfigure}[b]{0.49\textwidth}
         \centering
         \includegraphics[width=\textwidth]{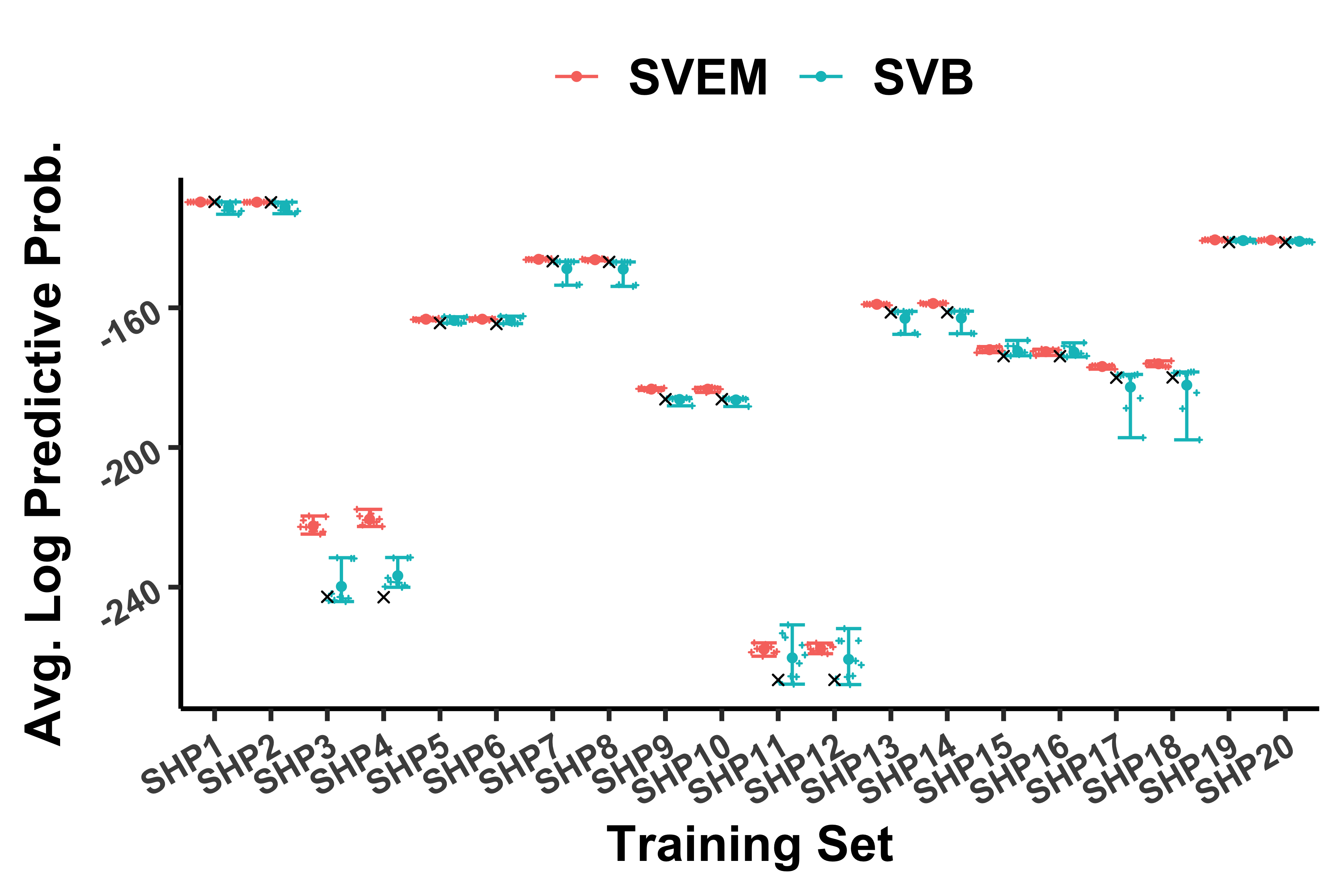}
         \caption{\textbf{Test set from SHP1: $\bm{y}^{(new)}_{Shp1}$.}}
         \label{lpp.png}
     \end{subfigure}
     \begin{subfigure}[b]{0.49\textwidth}
         \centering
         \includegraphics[width=\textwidth]{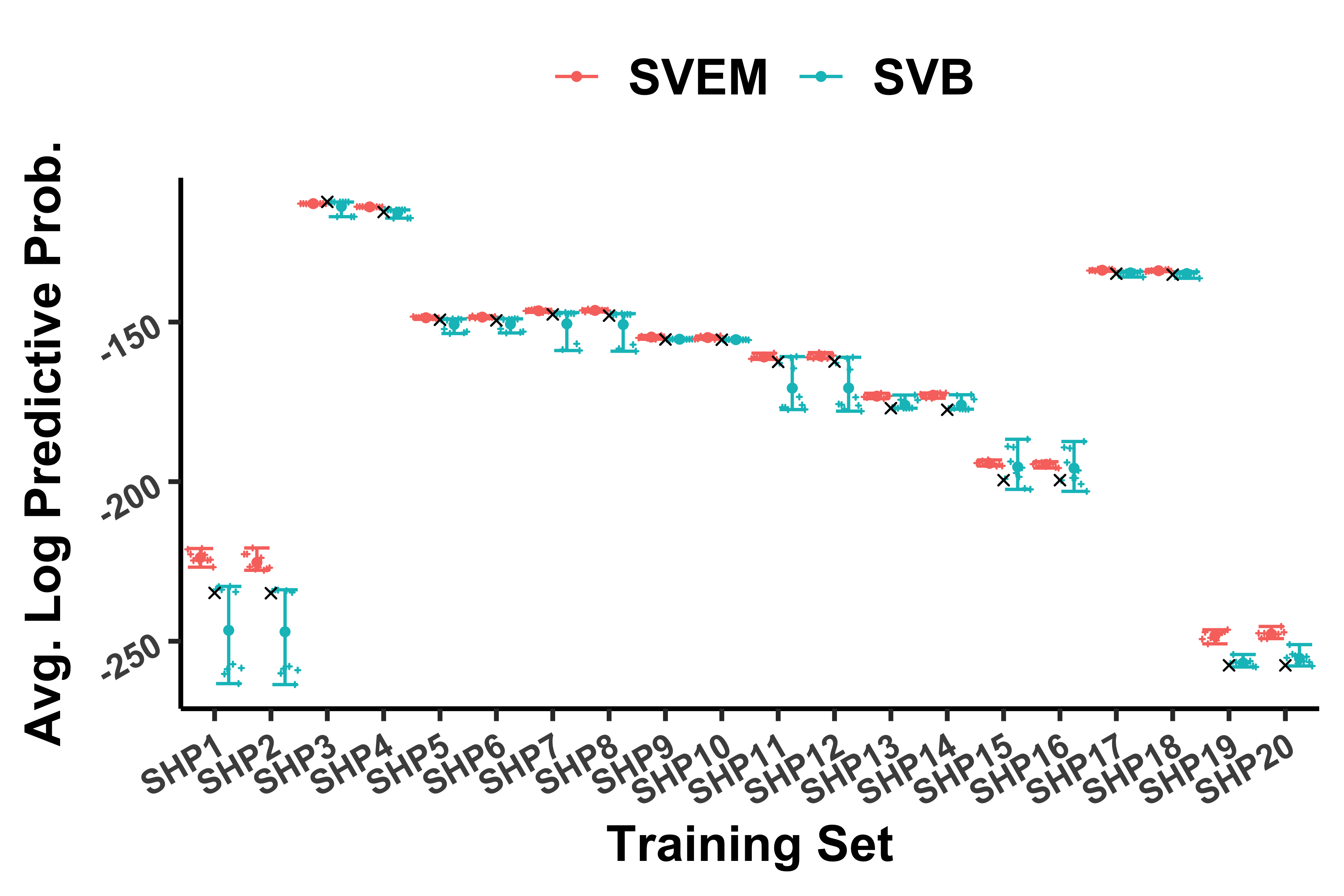}
         \caption{\textbf{Test set from SHP3: $\bm{y}^{(new)}_{Shp3}$.}}
         \label{pred_S3.png}
     \end{subfigure}
\caption{Log predictive probability averaged over 500 test sequences, using two point estimates: SVB in blue and SVEM in red. 
In panel (a), 
$ \ln p(\bm{y}^{(new)}_{Shp1} | \hat{\theta} ) $ uses $\hat{\theta}$ estimated from training sets of SHP1 to SHP20 (labeled on x-axis).
Every bar contains 10 approximated values of log predictive probability,
based on $\hat{\theta}$ trained from 10 initializations.
The mean and range of the 10 values are shown. 
Black cross dots represent $\ln p(\bm{y}^{(new)}_{Shp1} | \theta^{\ast} )$ under true parameters of each SHP, namely $\theta^{\ast}_{Shp1}$, ..., $\theta^{\ast}_{Shp20}$.
In panel (b) similar information is shown but with $\bm{y}^{(new)}_{Shp3} $ from the test set of SHP3.}
\label{fig:lpp}
\end{figure}

For each of the 20 SHPs, we used its test set (containing 500 sequences) as $\bm{y}^{(new)}$ (namely $\bm{y}^{(new)}_{Shp1}$ to $\bm{y}^{(new)}_{Shp20}$).
We used its training set (containing 4500 sequences) to generate 10 SVB estimates and 10 SVEM estimates based on the 10 random initializations. 
Figure \ref{fig:lpp} illustrates the log predictive probability using $\bm{y}^{(new)}_{Shp1}$
(Figure \ref{fig:lpp}(a)) and $\bm{y}^{(new)}_{Shp3}$ (Figure \ref{fig:lpp}(b)). The average over 500 test sequences is reported.
Results using $\bm{y}^{(new)}$ from other SHPs provide similar information and are not included.
Here in both panel (a) and (b), the x-axis indexes training sets of the 20 SHPs.
For each training set,
10 approximated log predictive probabilities are plotted in blue using the 10 SVB estimates, and the other 10 points are plotted in red using the 10 SVEM estimates.

In Figure \ref{fig:lpp}(a) when 
setting test sequences as $\bm{y}^{(new)}_{Shp1}$, 
the highest log predictive probability is obtained when the parameter estimate $\hat{\theta}$ (i.e., $\mathbb{E}_{ \lambda_\theta^{(t)} } \theta$ for SVB, $\theta^{(t)}$ for SVEM) is learned from SHP1.
That is, among all $\hat{\theta} $ obtained from 20 SHPs, log predictive probability $ \ln p(\bm{y}^{(new)}_{Shp1} | \bm{y}^{(1:N)}) \approx  \ln p(\bm{y}^{(new)}_{Shp1} | \hat{\theta} ) $ is maximized when $\hat{\theta} $ best approaches SHP1's true parameter $\theta^{\ast}_{Shp1}$.
The second highest log predictive probability is obtained when $\hat{\theta}$ is learned from SHP2, the ``twin'' of SHP1,
as expected due to our simulation design.
The lower values of $\ln p(\bm{y}^{(new)}_{Shp1} | \hat{\theta} ) $, for example when $\hat{\theta}$ is learned from SHP11,
suggest that SHP11 is more different from SHP1.
However, we should be cautious to what extent we can quantify such difference using log predictive probability (see our discussions in Section \ref{subsec:real}).
Similar to Figure \ref{fig:lpp}(a), Figure \ref{fig:lpp}(b) uses $\bm{y}^{(new)}_{Shp3}$ from the test set in SHP3,
and shows that the two highest log predictive probabilities are obtained with $\hat{\theta}$ learned from SHP3 and its ``twin'' SHP4.
In addition, Figure \ref{fig:lpp}(a) shows that SHP1/2 have greater similarity to SHP19/20 than to other SHPs,
and both SHP1/2 and SHP19/20 present the least similarity to SHP3/4 in Figure \ref{fig:lpp}(b).

Figure \ref{fig:lpp} illustrates that log predictive probability can be indicative of relative similarity 
among different RHP datasets.
We note that difference between the two log predictive probabilities can be approximated by the log likelihood ratio.
In addition,
Figure \ref{fig:lpp} echoes Figure \ref{fig:se} in that both SVB and SVEM provide good estimates to the likelihoods under $\theta^{\ast}_{Shp1}$, ..., $\theta^{\ast}_{Shp20}$ (labeled in black cross dots), whereas SVB exhibits less robustness to random initializations.

\subsection{Real Data Study}
\label{subsec:real}

The simulation study shows that through the GHSMM-RHP model, 
log predictive probability can be used to indicate the similarities among different RHP datasets.
In the real data study, we aim to use the log predictive probability to understand the biochemical properties of new RHP datasets in a predictable way.

\textit{\textbf{Data and Training Settings}}
The real data in \cite{jiang2020single} consists of seven synthesized RHP datasets, with different MMA/EHMA ratios 
to explore their varying proton transport rates. 
Specifically, the datasets are RHP50/20 (i.e., MMA/EHMA ratio is 50\% : 20\%), RHP70/0, RHP60/10, RHP35/35, RHP30/40, RHP40/30, and RHP55/15.
In each RHP dataset, 500 chains were held out and used as the test set, while the other 4500 chains were used as the training set. Each RHP chain is of length 130.
We used the same training settings as described in Section \ref{sec:simulation setup}.

\begin{figure}
\centering
     \begin{subfigure}[b]{0.49\textwidth}
         \centering
         \includegraphics[width=\textwidth]{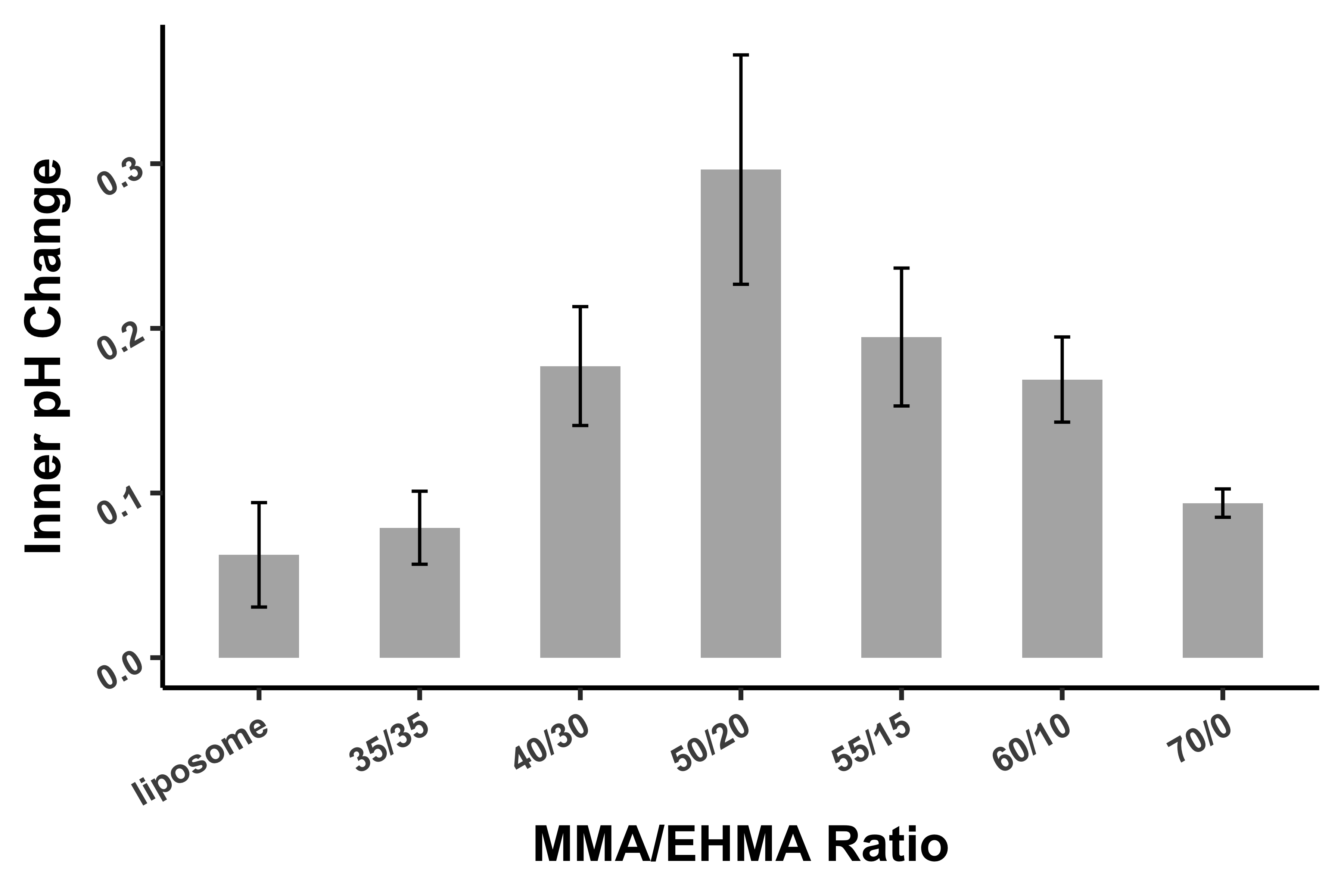}
         \caption{\textbf{Lab experimental results of RHPs.}}
         \label{pred_S3.png}
     \end{subfigure}
     \begin{subfigure}[b]{0.49\textwidth}
         \centering
         \includegraphics[width=\textwidth]{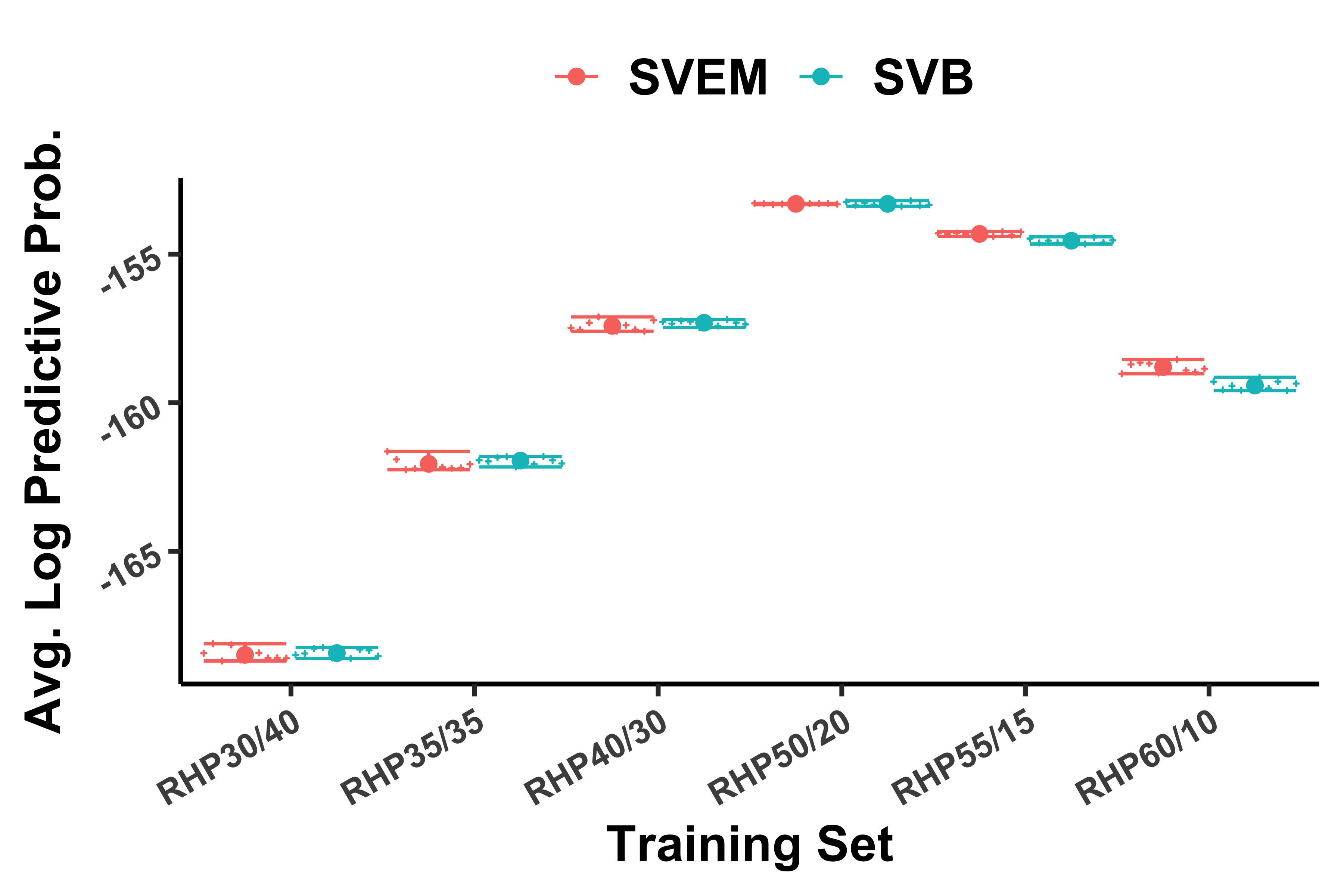}
         \caption{\textbf{GHSMM-RHP model results of RHPs.}}
         \label{lpp.png}
     \end{subfigure}
\caption{
Panel (a) adapts results from \cite{jiang2020single}, showing the proton transport performance of different RHPs measured by inner pH change (a higher measurement is better); error bars are of one standard deviation.
Panel (b) plots log predictive probabilities averaged over 500 test sequences from RHP50/20.
Training sets are labeled on the x-axis, with SVB estimates in blue and SVEM in red. 
Each bar contains 10 approximated values of log predictive probability, based on $\hat{\theta}$ trained from 10 initializations. The mean and range of the 10 values are shown.}
\label{fig:real}
\end{figure}

Figure \ref{fig:real}(a) shows the experimentally measured proton-transport performance of the RHPs with varying MMA/EHMA ratios \citep{jiang2020single}.
The assay was conducted on liposomes (sphere-shaped vesicles with the membrane composed of lipid bilayers) with a lower initial pH inside and a higher initial pH outside. The inner pH change for the liposome is a measure of protons that transport across the RHP-lipid bilayers. 
Four RHP datasets that promote the proton transport rate were reported.
Among them RHP50/20 has the best performance, followed by RHP55/15, RHP40/30, and RHP60/10. 
However, the 70/0, 35/35 ratios do not lead to distinct difference compared with the liposome sample, which suggests their lower proton transport efficiency.
It was argued that with the MMA/EHMA ratio deviating from 50/20, the changed overall hydrophilic-lipophilic balance could affect the interplay between polymers and the bilayers, and obstruct the RHP insertion.
Note that there was no experimental result for RHP30/40 so it is not shown in panel (a).

\textit{\textbf{Predictive Inference Results}}
We repeated the analysis in Section \ref{sec:lpp} and
applied the two stochastic variational methods to the real data.
By setting $\bm{y}^{(new)}$ to the test set of RHP50/20 which is the dataset of the highest inner pH change,
and computing log predictive probabilities based on $\hat{\theta}$ learned from other RHPs,
we hope to understand 
other RHPs' proton transport efficiency
in a predictable manner.

The approximated log predictive probabilities are plotted in Figure \ref{fig:real}(b).
The highest log predictive probability is obtained when $\hat{\theta}$ is learned from RHP50/20 itself.
The second highest is obtained when $\hat{\theta}$ is learned from RHP55/15, followed by RHP40/30 and RHP60/10.
Note that RHP70/0 is omitted from Figure \ref{fig:real}(b) due to its extremely low value of log predictive probability (mean at -484.8859 for SVEM, -374.0407 for SVB).

Measuring protein-like behaviors in lab experiments can help to validate the findings in our model. For the four functional RHP datasets described in \cite{jiang2020single}, 
the rankings by log predictive probability (shown in Figure \ref{fig:real}(b)) are consistent with the rankings by inner pH change (shown in Figure \ref{fig:real}(a)).
That is, RHP50/20 followed by RHP55/15, RHP40/30 and RHP60/10.
Here, RHP50/20 exhibits the best transport performance and the largest inner pH change reading. 
For RHPs showing high statistical similarity to RHP50/20, 
their log predictive probability serves as a good indicator of their relative proton transport performance.
That is, those RHPs of higher rankings by GHSMM-RHP are datasets of experimental interest for their potential transport efficiency.
However, as discussed in Section \ref{sec:lpp},
the RHPs of the lower rankings, namely RHP 70/0, RHP35/35, and RHP30/40, are indicative of their relative inactive proton transport functionality, and we need to be cautious to use their exact rank.
This is because there are many other biochemical properties that can not be measured by inner pH change.
The greater dissimilarity indicated by predictive probability between RHP datasets may not necessarily lead to greater difference in inner pH change.

More importantly, Figure \ref{fig:real} shows that
the GHSMM-RHP model can provide guidance to synthesizing RHPs with better potential functionalities.
For example, sequences of RHPs are easy to generate in computer simulations, while their biochemical properties are expensive to analyze in the lab.
To this end,
we could use log predictive probability to select RHP datasets from a wide range of compositional ratios that are most similar to a target RHP of desired properties (such as RHP50/20).
The RHPs selected by GHSMM-RHP then have the greatest potential to exhibit the targeted biochemical properties (such as high inner pH change), and are 
favorable to be further validated in costly wet lab experiments.

The GHSMM-RHP model can also be applied to other RHP studies with different monomer choices (other than MMA, EHMA, OEGMA, and SPMA) and composition ratios. Without testing each RHP variants in the wet lab, the current study presents a reliable approach to predict the performance of RHP systems of varying heterogeneity. Such a cost- and time-efficient endeavor holds great potential for facilitating the design of functional RHPs in other application fields.

\section{Discussions}
\label{sec:Conc}

Our model constraints are derived from the result of the sequential analysis in \cite{jiang2020single}, as well as our current empirical understanding of RHP systems.
Despite the limited domain information,
we believe the GHSMM-RHP model captures the critical connections between hidden segment structures and the RHP functionality to some extent.
Moreover, we can adjust model settings such as segment length constraints whenever there is new supporting evidence.

The similarity measure we used is built on the predictive inference in Bayesian statistics. In particular, the approximation to predictive probability comes down to approximated likelihood.
This relates to the log likelihood ratio statistics when comparing two RHP datasets using their log predictive probabilities. Other statistical tools related to likelihood ratio statistics may be explored for application in RHP study.

The GHSMM-RHP model not only holds potential for measuring the similarity of existing RHP datasets,
but can also be exploited to create new groups of RHP sequences.
This may be done by iteratively refining the current RHP dataset, removing dissimilar sequences and adding in new sequences of greater similarity, until some termination criterion is met. 
The newly generated RHP dataset may present novel monomer composition ratios, and offer insights into the target functionality.
GHSMM-RHP provides new ways of exploiting the rich heterogeneity as a key factor in designing functional RHPs.

We also empirically studied the computational performance of stochastic variational methods under Bayesian and frequentist frameworks. 
Our work takes an initial effort to compare the two methods in the same non-convex estimation-prediction problem, in which we carried out an extensive simulation study and a real data application.
There could be other hyper-parameters besides the four factors we tried (initializations, learning rates, batch size, minimum segment length) that affect the practical performance of algorithms, and more tuning efforts may be implemented.

We note that the two methods discussed in this paper are termed variational in the spirit of
using optimization for density estimation \citep{wainwright2008graphical, blei2017variational}. 
It has also been argued that EM is itself a mean-field variational method \citep{neal1998view, hoffman2013stochastic}.
As illustrated,
we provided the objective functions of Bayesian and frequentist variational methods in Section \ref{sec:Method} using the same variational principle based on KL divergence.
Both methods involve stochastic optimizations in the M step.
They justify the terminologies we used in this paper post hoc.
Similar ideas have been proposed to use only partial information from the entire dataset in the hope of scaling up the inference algorithm. 
Such efforts include incremental EM \citep{neal1998view}, first order EM \citep{salakhutdinov2003optimization}, online EM \citep{cappe2009line},
as well as \cite{hoffman2013stochastic} who introduced 
the first order optimization to variational methods under the Bayesian framework.

{\large\bf Supplementary Material}

More implementation details on the message-passing algorithms, derivation of the alternate optimizations, code and RHP datasets are available in the supplementary material.

{\large\bf Acknowledgments}

The study was supported by
DOD HDTRA1-19-1-0011 and DA ARO W911NF2110128.



%

\begingroup
    \setlength{\bibsep}{4.65pt} 
    \setstretch{1}
   \bibliographystyle{plainnat}
   \bibliography{Bibliography-MM-MC}
\endgroup



\end{document}